\documentclass[%
 reprint,
 amsmath,amssymb,
 aps,
]{revtex4-2}

\usepackage{graphicx}% Include figure files
\usepackage{dcolumn}% Align table columns on decimal point
\usepackage{bm, xcolor}
\usepackage{array, multirow, makecell, hhline}
\usepackage[hidelinks]{hyperref}% add hypertext capabilities
\usepackage[version=4]{mhchem}

\usepackage{subfig} 
\usepackage{verbatim}
\usepackage{graphics,setspace,float,enumitem}
\usepackage{commath,mathrsfs} % amsbsy, mathabx

\newcommand{\angstrom}{\mbox{\normalfont\AA}}
\newcommand{\Trace}[1]{\mathrm{Tr}\left[#1\right]}
\newcommand{\bra}[1]{\langle #1|}
\newcommand{\ket}[1]{|#1\rangle}

\newcommand{\DFTU}{DFT+$U$}
\newcommand{\DFTUJ}{DFT+$U$+$J$}
\newcommand{\DFTUpJp}{DFT+$U$(+$J$)}

\newcommand{\Edft}{E_{\text{DFT}}}
\newcommand{\Euj}{E_{U/J}}
\newcommand{\Eu}{E_{U}}
\newcommand{\Edftuj}{E_{\text{\DFTUJ{}}}}
\newcommand{\Ehub}{E_{\text{Hub}}}
\newcommand{\Edc}{E_{\text{dc}}}

\begin{document}

\preprint{APS/123-QED}

\title{High-throughput determination of Hubbard $U$ and Hund $J$ values for transition metal oxides via the linear response formalism}% Force line breaks with \\

\author{Guy C. Moore\textsuperscript{1,2}}
\author{Matthew K. Horton\textsuperscript{1,2}}
\author{Alexander M. Ganose\textsuperscript{3}}
\author{Martin Siron\textsuperscript{1,2}}
\author{Edward Linscott\textsuperscript{4}}
\author{David D. O'Regan\textsuperscript{5}}
\author{Kristin A. Persson\textsuperscript{1,6}}
\affiliation{
    \textsuperscript{1}Department of Materials Science and Engineering, University of California Berkeley, Berkeley, CA 94720, \\
    \textsuperscript{2}Materials Science Division, Lawrence Berkeley National Laboratory, Berkeley, CA 94720, USA, \\
    \textsuperscript{3}Energy Technologies Area, Lawrence Berkeley National Laboratory, Berkeley, CA 94720, USA, \\
    \textsuperscript{4}Theory and Simulations of Materials (THEOS), and National Centre for Computational Design and Discovery of Novel Materials (MARVEL), \'Ecole Polytechnique F\'ed\'erale de Lausanne, 1015 Lausanne, Switzerland, \\
    \textsuperscript{5}School of Physics, SFI AMBER Centre and CRANN Institute, Trinity College Dublin, The University of Dublin, Ireland, \\
    \textsuperscript{6}Molecular Foundry, Lawrence Berkeley National Laboratory, Berkeley, CA 94720, USA
}

\date{\today}% It is always \today, today,

\begin{abstract}
\DFTU{} provides a convenient, cost-effective correction for the self-interaction error (SIE) that arises when describing correlated electronic states using conventional approximate density functional theory (DFT). The success of a \DFTUpJp{} calculation hinges on the accurate determination of its Hubbard $U$ and Hund's $J$ parameters, and the linear response (LR) methodology has proven to be computationally effective and accurate for calculating these parameters. This study provides a high-throughput computational analysis of the $U$ and $J$ values for  transition metal $d$-electron states in a representative set of over 2000 magnetic transition metal oxides (TMOs), providing a frame of reference for researchers who use DFT$+U$ to study transition metal oxides. In order to perform this high-throughput study, an \texttt{atomate} workflow is developed for calculating $U$ and $J$ values automatically on massively parallel supercomputing architectures. To demonstrate an application of this workflow, the spin-canting magnetic structure and unit cell parameters of the multiferroic olivine \ce{LiNiPO4} are calculated using the computed Hubbard $U$ and Hund $J$ values for Ni-$d$ and O-$p$ states, and are compared with experiment. Both the Ni-$d$ $U$ and $J$ corrections have a strong effect on the Ni-moment canting angle. Additionally, including a O-$p$ $U$ value results in a significantly improved agreement between the computed lattice parameters and experiment.
\begin{comment}
\begin{description}
\item[Usage]
Secondary publications and information retrieval purposes.
\item[Structure]
You may use the \texttt{description} environment to structure your abstract;
use the optional argument of the \verb+\item+ command to give the category of each item. 
\end{description}
\end{comment}
\end{abstract}

\keywords{linear response, Hubbard U, Hund J}  % Use showkeys class option if keyword display desired

\maketitle

% \tableofcontents

\section{Introduction}
Density functional theory (DFT) is a workhorse of computational materials science. However, the proper treatment of electronic exchange and correlation within the framework of DFT is a long-standing challenge \cite{correlated-electrons-book-2012}.
%The Born–Oppenheimer approximation offers a simplification to the multi-body problem of interacting electrons and nuclei.
% It is computationally infeasible to take into account the interactions between all electrons in a material, which is one reason why Hartree-Fock (HF) theory has accelerated the computational speed of ab initio calculations, and their parallelizability for modern computational architectures. 
Local density approximation (LDA) and generalized gradient (GGA) \cite{gga-paper-OG} functionals were developed to add exchange-correlation (XC) contributions to the energy functional within the Kohn–Sham (KS) formalism \cite{ks-OG}. However, numerous studies have shown that these XC functionals have an associated self-interaction error (SIE) \cite{PhysRevB.71.035105, LinscottEtAl, correlated-electrons-book-2012}. This shortcoming ultimately derives from the difficulty in quantifying exact exchange and correlation effects, without solving the many-body Schr\"odinger equation, using only density-based approximations.

% ``Meta-GGA" functionals (such as SCAN \cite{PhysRevLett.115.036402}) theoretically reduce the severity of these errors by including kinetic energy terms in the exchange-correlation corrections to the free energy functional \cite{PhysRevLett.115.036402, corr-el-book-2012-ch4}. However, meta-GGA DFT implementations often suffer from a higher computational cost compared to their GGA counterparts. In addition, when it comes to magnetic properties, meta-GGA functionals can produce results that are further from experiment compared with their more conventional local or semi-local DFT counterparts \cite{shortcoming-metaGGA}. 

% Hybrid functional reference: \cite{flores-accuracy-2018}
% Coauthors Akash Bajaj, Jon P. Janet, and Heather J. Kulik 

Over the past couple of decades, DFT$+U$ has found favor as a method that strikes a reasonable balance between accuracy and computational cost, making it particularly suitable for high-throughput computation \cite{Anisimov1991,Anisimov1993,Anisimov1997,Pickett1998,corr-el-book-2012-ch4}. DFT$+U$ functionals add a correction to the conventional XC functional to account for the Coulombic interaction between localized electrons \cite{PhysRevB.71.035105, kulik-density-2006}. 
In more recent studies, various researchers have explored extensions of \DFTU{} with the goal of further correcting for static correlation effects and delocalization errors \cite{himmetoglu-hubbard-corrected-2014, bajaj-communication-2017, bajaj-molecular-2021}. 

One drawback to DFT$+U$ type functionals is that one must first determine its associated parameters, the Hubbard $U$
and Hund $J$, and possibly also inter-site electronic interactions denoted as ``+$V$" \cite{campo_jr_extended_2010, tancogne-dejean_parameter-free_2020,lee_first-principles_2020}. The results of a \DFTU{} calculation can quantitatively and even qualitatively change depending on these parameters, and so obtaining reliable values is of paramount importance.

This is as true for the Hund $J$ as it is for the Hubbard $U$, 
as we now explain.
In this work, we primarily focus on the simplified rotationally invariant \DFTU{} functional that has become very prominent since its introduction in  Ref.~\onlinecite{Dudarev-OG}. In this 
functional, the Hubbard $U$ and Hund $J$ are grouped in  single effective Hubbard $U$ parameter, $U_{\text{eff}}$, defined as $U_{\text{eff}} = U - J$. This  formalism assumes a spherical symmetry that results in the exclusion from the correction of the on-site exchange between opposite-spin electrons~\cite{Dudarev-OG, GGAplusUJ-LiNiPO4, corr-el-book-2012-ch4}. 
Notwithstanding, the reduction in the effective parameter 
by $J$ can be significant.

While the aforementioned approximation may seem
more justifiable for systems with no magnetic order, in the case of magnetic systems it results in a lost opportunity to use the Hund $J$ to beneficially enhance the spin moments in simulated broken-symmetry ground states.
Moreover, when we move to non-collinear magnetism, the spin texture of materials is particularly sensitive to  screening interactions between spin channels \cite{bultmark-multipole-2009, dudarev-ncl-UO2-2019, GGAplusUJ-LiNiPO4}.
In fact, magnetic exchange constants can be derived from the extended Hubbard model and estimated as ratios between $U$ and $J$ values \cite{streltsov-orbital-phys-2017}. The famous Hubbard model provides a simplified framework on which to explain the rich physics of correlated transition metal compounds \cite{streltsov-orbital-phys-2017}.
Additionally, it has been shown that the Hund $J$ term is important for describing important physical phenomena, such as Jahn-Teller distortions \cite{PhysRevB.92.085151, streltsov-orbital-phys-2017}, emergent intra-atomic exchange, and the Kondo effect \cite{doi:10.1146/annurev-conmatphys-020911-125045, herper_combining_2017}. Therefore, the introduction of explicit unlike-spin exchange corrections beyond simplified rotationally invariant \DFTU{} is clearly of interest, and this  requires the treatment of the 
Hund $J$ on the same footing as the Hubbard $U$.

\subsection{Strategies for determining Hubbard parameters}
A common approach for determining Hubbard $U$ values is to tune them such that some desired result --- for example, the \DFTU{} band gap, or a formation energy --- matches its experimental value, or a value obtained via more accurate and computationally expensive beyond-DFT methods \cite{jain-formation-2011,yu-machine-2020}. There are several problems with this strategy. Firstly, it is not systematic: just because one result (e.g., the band gap) now matches experiment, this does not guarantee the same will be true for other observables (e.g., local magnetic moments). Indeed, there a multitude of reasons why DFT may not match experiment, and it is wrong to rely on Hubbard corrections to correct for errors that do not arise from self-interaction \cite{Albers-2009}. Secondly, this strategy is not predictive: it relies on the existence of experimental/beyond-DFT data. This makes it particularly ill-suited to the prediction of novel materials and high-throughput studies.

%\textcolor{red}{I'm not quite sure what to do with the following paragraph. It also conflates Hund's coupling and FSL} 

 %Furthermore, on a theoretical basis, it is possible to extend the motivation of Hubbard $U$ to Hund $J$ as a correction to enforce piece-wise linearity of the energy of a system versus on-site occupation numbers \cite{bajaj-communication-2017}. On this subject, Bajaj and others have explored the importance of Hund $J$ in maintaining the ``fractional spin line" (FSL) along spin degrees of freedom \cite{bajaj-communication-2017}.

Yet another difficulty that arises is the lack of transferability of Hubbard and Hund's parameters. The conventional wisdom surrounding on-site corrections tends to reinforce the notion that localization equals correlation. Therefore, +$U$ corrections are applied to states determined by the orbital geometry (e.g., $d$ and $f$ orbitals). It is possible, using spectroscopy, to estimate Slater integrals over the Coulomb operator \cite{Dudarev-OG, elfimov-magnetizing-2007} which, in turn, can be expressed in terms of $U$ and $J$ values \cite{Liechtenstein-OG, corr-el-book-2012-ch4}. While this connection is physically motivated, localized states do not encompass all of the levels of correlation effects that are neglected by the specific DFT functional \cite{bajaj-communication-2017}. This perspective of $U$ and $J$ values as functional-specific, and not universal quantities, expands the definition of $U$ and $J$ from their initial inspiration from the Hubbard model, which treats $U$ and $J$ as intrinsic atomic properties. Indeed, it has been repeatedly shown that these parameters $U$ and $J$ are in fact very sensitive to the local chemical environment \cite{cococcioni-ceder-ldau-redox}.
% \textcolor{red}{Not sure who best to cite here, especially given we haven't introduced LR yet}.
Even the specific pseudopotentials (PPs) \cite{LinscottEtAl} or the specific site occupation projection scheme \cite{wang_local_2016} have a significant effect on the computed Hubbard $U$ values. The end result is that $U$ values
(and by extension the albeit normally less 
environment-sensitive Hund's $J$ values) are not transferable: 
they cannot be tabulated, and must always be determined on a 
case-by-case basis.

% \textcolor{red}{DUMPING THIS HERE, NEEDS TO BE INTEGRATED INTO SURROUNDING TEXT}
Having explored the numerous on-site corrections, and the drawbacks of fitting these parameters to experiment or beyond-DFT results, we will motivate the importance of computing Hubbard $U$ and Hund $J$ values within the DFT framework. Two primary methods for calculating Hubbard $U$ values in a self-contained fashion within DFT are the constrained random phase approximation (cRPA) \cite{corr-el-book-2019-ch2, vaugier-hubbard-2012}, and the linear response (LR) analysis of the constrained XC functional \cite{corr-el-book-2012-ch4, PhysRevB.71.035105}. In this study, we focus on the LR method due to its lower computational cost compared to existing cRPA methods, which are not yet appropriate for high-throughput applications. We also explore the effects on magnetic materials that exhibit a rich variety of noncollinear spin configurations, exemplified through the spin canting structure that was experimentally observed in olivine \ce{LiNiPO4} \cite{exp-LiNiPO4}.

The linear response method, as introduced for practical use 
by Coccocioni and coworkers \cite{PhysRevB.71.035105}, is founded on the idea that SIE can be related to the behaviour of the total energy as a function of the total occupation \cite{Cohen2008a}. The energy ought to be piece-wise linear with respect to total site occupation numbers, but in fact for semi-local DFT XC functionals, the energy derivatives are erroneously continuous. Cococcioni and co-workers illustrated that the $+U$ correction can be interpreted as something that counteracts this erroneous curvature, 
locally for sub-spaces (where the interpretation becomes approximate). Crucially, the magnitude of the curvature can be directly measured from a DFT linear response calculation, allowing the value of $U$ to be determined accordingly. Unlike empirical fitting, this approach is (a) systematic, because the value of $U$ is derived directly as a measure of the underlying SIE present in the DFT calculation, and (b) it is predictive, because it only requires DFT calculations to extract the Hubbard parameters, and not experimental or beyond-DFT results.

%\textcolor{red}{Not sure where the following belongs}

%Cococcioni has reported that USPPs result in longed-ranged interactions in the linear response \cite{corr-el-book-2012-ch4}, however, we are unable to find a demonstration of this. Within the scope of this study, it was not readily possible to compare the result of using USPPs, as performing \DFTU{} calculations using USPPs is not currently implemented in VASP.

\subsection{Paper outline}
The Materials Project is a web-based database that contains computed information on a vast range of materials, both known and predicted \cite{Jain2013}. Among the various computational results it presents are Hubbard parameters $U_\mathrm{eff}$. However, these current default $U_{\text{eff}}$ values were obtained by fitting \DFTU{} energies to experimental formation energies for a selected number of redox reactions \cite{mp-uvals, cococcioni-ceder-ldau-redox}. This paper aims to replace these values with ones computed using linear response. In order to achieve this, we present a unified framework for computing on-site Hubbard and Hund's corrections in a fully parallelized and automated computational workflow (which will be introduced in Section \ref{sec:methods}).
%We presume that values calculated using this framework will be useful for obtaining more accurate defect formation energies, particularly in the study of oxygen vacancies \cite{scf-u-SrMnO3}.
Using this workflow, we performed a high-throughput calculation of $U$ and $J$ values for a set of over two thousand transition-metal-containing compounds. This provides us with a novel, big-picture point-of-reference for the sensitivity of $U$ and $J$ across a wide range of systems of varying chemistries and local chemical environments (Sections~\ref{sec:periodic_table} and \ref{sec:MnFeNiO}).
% \textcolor{red}{Cut/move the following?} We also explore the possible trade-offs of using a PAW projection scheme, in addition to different pseudopotentials in the Supplementary Information. We've observed improved supercell scaling compared to other studies \cite{PhysRevB.71.035105, bennett_systematic_2019} that use different DFT implementations. However, the calculated $U$ and $J$ values have a larger uncertainty compared to other studies \cite{LinscottEtAl}.
 We then explore the effects of these Hubbard corrections on magnetic materials that exhibit a rich variety of noncollinear spin configurations, exemplified through the spin canting structure of olivine \ce{LiNiPO4} (Section~\ref{sec:LiNiPO4}).

\section{Methods}
\label{sec:methods}
\subsection{The Hubbard functional}

The Hubbard functional is a corrective functional, in the sense that it involves adding a corrective term $\Ehub{} - \Edc{}$ on top of some base functional $\Edft{}$ (typically a local or semi-local functional), resulting in a total energy functional
\begin{align}
  & \Edftuj{} \left[\rho,  \left\{ \bm n^\sigma_\gamma \right\} \right] \nonumber \\
  & \qquad = \Edft{} \left[\rho \right]  \nonumber \\
  & \qquad \quad + \Ehub{} \left[\left\{ \bm n^\sigma_\gamma \right\}\right]
  - \Edc{} \left[\left\{ n^\sigma_\gamma \right\}\right] \nonumber \\
  & \qquad = \Edft{} \left[\rho \right]
  + \Euj{} \left[\left\{ \bm n^\sigma_\gamma \right\}\right]
\label{eqn:General_DFT+U_energy}
\end{align}
The $(\bm n_\gamma ^\sigma)_{mm'} = \bra{\varphi_{\gamma m}} \hat \rho^{\sigma} \ket{\varphi_{\gamma m'}}$ are matrices that represent the projection of the (spin-dependent) density operator onto Hubbard subspaces (indexed $\gamma $) defined by some set of orbitals $\ket{\varphi_{\gamma m}}$. These orbitals are typically atom-centred, fixed, spin-independent, localised, and orthonormal, often corresponding to the $3d$ or $4f$ subshell of a transition metal or lanthanide. The $n_\gamma^\sigma$ occupation numbers are the corresponding traces of $\bm n_\gamma^\sigma$ matrices. The \DFTU{} correction of Equation \ref{eqn:General_DFT+U_energy} adds a convex energy penalty to fractional occupations of these orbitals that in principle can counterbalance the SIE present in these Hubbard subspaces.

% \textcolor{red}{FILL OUT AND EXPLAIN EACH METHOD: (a) Liechtenstein (b) Himmetoglu (c) Dudarev}

In the following paragraphs, we will provide a summary of some of the most well known formulations of \DFTUpJp{}. We note that that because we are interested in the fully localized limit (FLL), we will not discuss extensions of \DFTUJ{} to metallic systems, which employs an ``around mean field'' (AFM) methodology \cite{corr-el-book-2012-ch4}. 

Starting from \DFTUJ{} implementations of the highest complexity, and moving forward through increasing levels of simplification, we introduce the rotationally invariant implementation proposed by Liechtenstein et al. \cite{Liechtenstein-OG}. Within this flavor of \DFTUJ{}, $\Ehub{}$ and $\Edc{}$ take the following form
\begin{align}
  \Ehub{}
  &= \frac{1}{2} \sum_{\{m\},\gamma,\sigma} \mathcal U({m,m',m'',m'''}) (n^\sigma_\gamma)_{mm'} (n^\sigma_\gamma)_{m''m'''} \nonumber \\
  &- \frac{1}{2} \sum_{\{m\},\gamma,\sigma} \mathcal U({m,m'',m''',m'}) (n^\sigma_\gamma)_{mm'} (n^\sigma_\gamma)_{m''m'''}
\label{eqn:liech_hub_energy} \\
  \Edc{}
  &= \sum_{\gamma}  \frac{U_\gamma}{2} n_\gamma \left(n_\gamma - 1\right) 
  + \sum_{\gamma,\sigma} \frac{J_\gamma}{2} n^\sigma_\gamma \left(n^\sigma_\gamma - 1\right) ,
\label{eqn:liech_dc_energy}
\end{align}
where $\mathcal U$ contains the Coulomb integrals projected on the orbital basis, indicated by the associated $\{ m \}$ quantum numbers \cite{corr-el-book-2012-ch4, himmetoglu-hubbard-corrected-2014}. This correction is parameterized by both Hubbard $U_\gamma$ and Hund $J_\gamma$ coupling constants through the double-counting energy contribution, $\Edc{}$.

Simplified versions of Equations \ref{eqn:liech_hub_energy} \& \ref{eqn:liech_dc_energy} were proposed by Dudarev et al. \cite{Dudarev-OG}, and later by Himmetoglu and coworkers \cite{Himmetoglu2011a}, which approximate $\mathcal U$ using Slater integrals, which can be parameterized through $U$ and $J$ values. There are many helpful explanations for this approximation, such as those summarized in Refs.~ \onlinecite{corr-el-book-2012-ch4, himmetoglu-hubbard-corrected-2014}. 

In the spirit of following increasing levels of simplification, we will start with the Himmetoglu implementation \cite{Himmetoglu2011a}, inspired by the work of Solovyev et al. \cite{Solovyev1994a}. Using the Slater integral parameterization of $U$ and $J$, it is possible to approximate and
simplify $\Euj{}$ from Equations \ref{eqn:liech_hub_energy} \& \ref{eqn:liech_dc_energy} into the following
\begin{align}
  & \Euj{} = \Ehub{} - \Edc{} = \nonumber \\
  & \qquad \sum_{\gamma \sigma} \frac{U_\gamma - J_\gamma}{2}\Trace{\bm n_\gamma ^\sigma(1- \bm n_\gamma ^\sigma)}
  + \sum_{\gamma \sigma}\frac{J_\gamma }{2}\Trace{\bm n_\gamma ^\sigma \bm n_\gamma ^{-\sigma}} .
\label{eqn:himm_uj_energy}
\end{align}
A well known further simplification of Equation \ref{eqn:himm_uj_energy}, 
notwithstanding that it substantially pre-dated the latter, is the formulation of \DFTU{} put forth by Dudarev et al. \cite{Dudarev-OG} 
and given by
\begin{align}
  & \Eu{} = \Ehub{} - \Edc{} = 
  \sum_{\gamma \sigma} \frac{U^\text{eff}_\gamma }{2}\Trace{\bm n_\gamma ^\sigma(1- \bm n_\gamma ^\sigma)}.
\label{eqn:dudarev_u_energy}
\end{align}
As discussed in the Introduction, this approximation arises by assuming spherical symmetry of the Coulomb interactions, $\mathcal U$ \cite{Himmetoglu2011a, corr-el-book-2012-ch4, himmetoglu-hubbard-corrected-2014}. Within the simplified Dudarev \DFTU{}, Equation \ref{eqn:dudarev_u_energy}, it has been demonstrated that the effective Hubbard $U$ becomes $U^\text{eff}_\gamma = U_\gamma - J_\gamma$ \cite{Dudarev-OG, corr-el-book-2012-ch4, himmetoglu-hubbard-corrected-2014}.

%%%%%%%%
%%%%%%%%
\begin{comment}
$E_{\mathrm{DFT+}U} = E_\mathrm{DFT} + E_U$.
The corrective term is given by
%
\begin{align}
E_{U}[n_\gamma ^\sigma] = \sum_{\gamma \sigma} \frac{U_\gamma }{2}\Trace{n_\gamma ^\sigma(1-n_\gamma ^\sigma)}
\label{eqn:DFT+U_energy}
\end{align}
%
The $+J$ extension involves adding a second correction to the total energy,
% 
\begin{align}
E_{J}[n_\gamma ^\sigma] = \sum_{\gamma \sigma}\frac{J_\gamma }{2}\Trace{n_\gamma ^\sigma n_\gamma ^{-\sigma}},
\label{eqn:DFT+J_energy}
\end{align}
\end{comment}
%%%%%%%%
%%%%%%%%

\subsection{Hubbard $U$ and Hund's $J$ spin polarized linear response}
\label{sec:lr_theory}
In the linear-response approach, one measures the 
supposedly erroneous curvature in the total energy as a function of the subspace occupancy, and then chooses a value $U$ that counterbalances the observed curvature. Computing the energy curvature as a function of the subspace occupancy is usually impractical, so instead one transforms the curvature of the energy-versus-site occupancies $n_\gamma$  into a curvature with respect to the magnitude $v_\gamma$ of an on-site potential $\hat v_\gamma = \sum_{mm'} v_\gamma \ket{\varphi_{\gamma m}}\bra{\varphi_{\gamma m'}}$. The energy functional is then given by
\begin{align}
E[\{ v_\gamma\}] =  \text{min}_{\rho(\bm{r})} \left\{ E[\rho(\bm{r})] + \sum_\gamma v_\gamma n_\gamma \right\}
\end{align}
from which one computes the response matrices
\begin{align}
\chi_{\gamma \gamma'} &= \frac{\partial n_\gamma}{\partial v_\gamma'}. \label{eq:resp_general}
\end{align}
Thus far we have used a general index ``$\gamma$" to represent each site. Conventionally, this index refers purely to the atom $\gamma$ on which the Hubbard site is centered. In this case, the Hubbard parameter for that subspace is given by
\begin{align}
U_\gamma &= \left( \chi_0^{-1} - \chi^{-1} \right)_{\gamma\gamma}
\label{eq:U_conventional}
\end{align}
where $\chi$ and $\chi_0$ are the interacting, (or self-consistent) and non-interacting (or non-self consistent) response matrices \cite{corr-el-book-2012-ch4,PhysRevB.71.035105}.

The above strategy does not delineate between spin channels: during the linear-response calculations the spin-up and spin-down channels are perturbed simultaneously by the same amount, i.e., $v^\uparrow_\gamma = v^\downarrow_\gamma$ and we only observe the change in total occupancy $n_\gamma = n^\uparrow_\gamma + n^\downarrow_\gamma$. If we want to calculate $J$, one must instead consider the spin-dependent perturbation
\begin{align}
    \hat v^\sigma_\gamma =
    \begin{cases}
        + \sum_{mm'} v_\gamma \ket{\varphi_{\gamma m}}\bra{\varphi_{\gamma m'}} & \sigma = \uparrow \\
        - \sum_{mm'} v_\gamma \ket{\varphi_{\gamma m}}\bra{\varphi_{\gamma m'}} & \sigma = \downarrow
    \end{cases}
    \label{eq:beta_perturbation}
\end{align}
and then construct a second set of response matrices which then relate to $J$ in a completely parallel approach to the calculation of $U$ in \ref{eq:U_conventional}.

A separate but ultimately equivalent strategy is to treat the spin channels separately \cite{LinscottEtAl, lambert-dftuj-2021}. In this case a  general index runs over both the atom index $\gamma = \{1,...,N\}$ and also the two spin channels $\sigma = \{\uparrow, \downarrow\}$. In this case the response matrices of Equation~\ref{eq:resp_general} become rank-four tensors, i.e.,
\begin{align}
\chi^{\sigma \sigma'}_{\gamma\gamma'} &= \frac{\partial n^\sigma_{\gamma}}{\partial v^{\sigma'}_{\gamma'}}.
\end{align}
and now the equivalent of Equation~\ref{eq:U_conventional} is
\begin{align}
f^{\sigma\sigma'}_{\gamma\gamma} &= \left( \chi_0^{-1} - \chi^{-1} \right)^{\sigma\sigma'}_{\gamma\gamma} \label{eq:fmatrix}
\end{align}
where now we must now prescribe how to map the $2\times2$ matrix $f^{\sigma\sigma'}_{\gamma\gamma}$ to the scalar parameters $U_\gamma = G_U(f^{\sigma\sigma'}_{\gamma\gamma})$ and $J_\gamma = G_J(f^{\sigma\sigma'}_{\gamma\gamma})$. Possible definitions for these mappings $G_U$ and $G_J$ are motivated and explored in detail in Ref.~\onlinecite{LinscottEtAl}, but the end result is the following: there are two possible approaches. In the first approach one can define this mapping in order to recover the $U_\gamma$ and $J_\gamma$ that one would obtain using the conventional spin-agnostic approach of Equations~\ref{eq:U_conventional} and \ref{eq:beta_perturbation}. We will hereafter refer to this as the ``conventional" strategy (in the language of Ref.~\onlinecite{LinscottEtAl} this is the ``scaled" approach). In the second approach one can define the mapping to impose the condition that the local magnetic moment (local occupation) is  held fixed during the perturbation while calculating the Hubbard (Hund's) parameter, specifically by means
of the the equations rather than in the
explicit sense of  fixing these quantities
using constrained DFT. 
We will refer to this as the ``constrained" approach (the ``simple" approach in Ref.~\onlinecite{LinscottEtAl}). Throughout this work, unless otherwise stated, we will use the conventional strategy.

% \textcolor{red}{Could discuss the strengths of spin-resolved LR now, but alternatively don't want to distract (and create targets for the reviewers)}

\subsection{Implementation of linear response within a high-throughput workflow}

The linear response method was implemented as a workflow within the high-throughput \texttt{atomate} framework \cite{atomate-paper}. The workflow allows the user to compute Hubbard $U$ and Hund $J$ values using either a spin-polarized or a non-spin-polarized response. In addition to screening between spin channels, the implementation provides the straightforward extension to multiple levels of screening, including inter-site and inter-spin-channel responses \cite{LinscottEtAl}. A more detailed explanation of how these screening matrices are computed is provided in Appendix \ref{sec:inversion_schemes}.

All of the individual calculations within this workflow were performed with VASP (Vienna ab initio Simulation Package) \cite{hafner_vienna_1997}, a plane-wave DFT code. The PBE exchange-correlation functional was used throughout as the base functional \cite{Perdew1996a}. Unless otherwise stated we use PAW PBE pseudopotentials (PPs), which are the default PPs for the \texttt{pymatgen} input sets for VASP \cite{pymatgen-github}. 
In this regard, our work supplements the high-throughput work of Bennett et al. \cite{bennett_systematic_2019} where ultrasoft pseudopotentials (USPPs) were used to reduce computational cost in high-throughput computations \cite{bennett_systematic_2019}, 
mirroring early foundational studies on the linear response method  \cite{PhysRevB.71.035105, campo_jr_extended_2010}.

We have used an automatic $k$-point generation scheme that uses 50 $k$-points per reciprocal angstrom, and a cutoff energy of 520 eV. The full set of input parameters can be found in the \texttt{HubbardHundLinRespSet} in the \texttt{atomate} repository \cite{atomate-github}, and the derived VASP input sets in the \texttt{pymatgen} repository  \cite{pymatgen-github}. For the linear response analysis, the on-site applied potential $v_{I\sigma}$ range was from $-0.2$ eV to $+0.2$ eV ($-0.05$ eV to $+0.05$ eV for the periodic table data set) sampled at nine points at uniform intervals.

\section{Results}

Hubbard $U$ and Hund $J$ values were calculated for over two thousand transition metal oxides using the linear response workflow implemented in \texttt{atomate}. The majority of the calculations corresponded to materials containing Mn-$d$, Fe-$d$, and/or Ni-$d$ species. All the systems studied were previously predicted by Ref.~\onlinecite{horton-high-throughput-2019} to have a collinear magnetic ground-state using a separate high-throughput workflow. That work used the empirical Hubbard $U$ values reported on the Materials Project.

In addition, a representative set of O-$p$ responses were calculated and analyzed. It is less common to include Hubbard corrections to oxygen 2$p$ states. However, an appreciable number of studies have shown how O-$p$ on-site corrections have improved the agreement with experimentally measured bond lengths between oxygen and transition metal species \cite{goh-effects-2017, LinscottEtAl, WO3-U-Op, plata-communication-2012, kuang-ab-2014}. It is perhaps less intuitive to apply spin-polarized Hund $J$ parameters to oxygen sites, because O-$p$ states are conventionally not included in effective models for magnetism. 
%
% However, in a previous study, some of us have argued that these O-$p$ $J$ values are justified, and can have significant effects on the resulting properties \cite{LinscottEtAl}. 
However, while oxygen atoms do not develop magnetic moments, early studies have demonstrated theoretically and computationally that O-$p$ states mediate the antiferromagnetic superexchange interaction in transition metal oxides, such as MnO \cite{streltsov-orbital-phys-2017, kramers-interaction-1934, anderson-antiferromagnetism-1950}. 

% \textcolor{red}{Discuss about response, stress that the oxygens don't develop a magnetic moment etc} 

% Therefore, it would be interesting to explore how the $J$ value on O-$p$ states could help to improve the quantification of effects such as superexchange.

\begin{figure*}
    \centering
    \subfloat[\centering Hubbard $U$ Periodic Table]{{
    \label{fig:ptable_u}
    \includegraphics[width=0.90\linewidth]{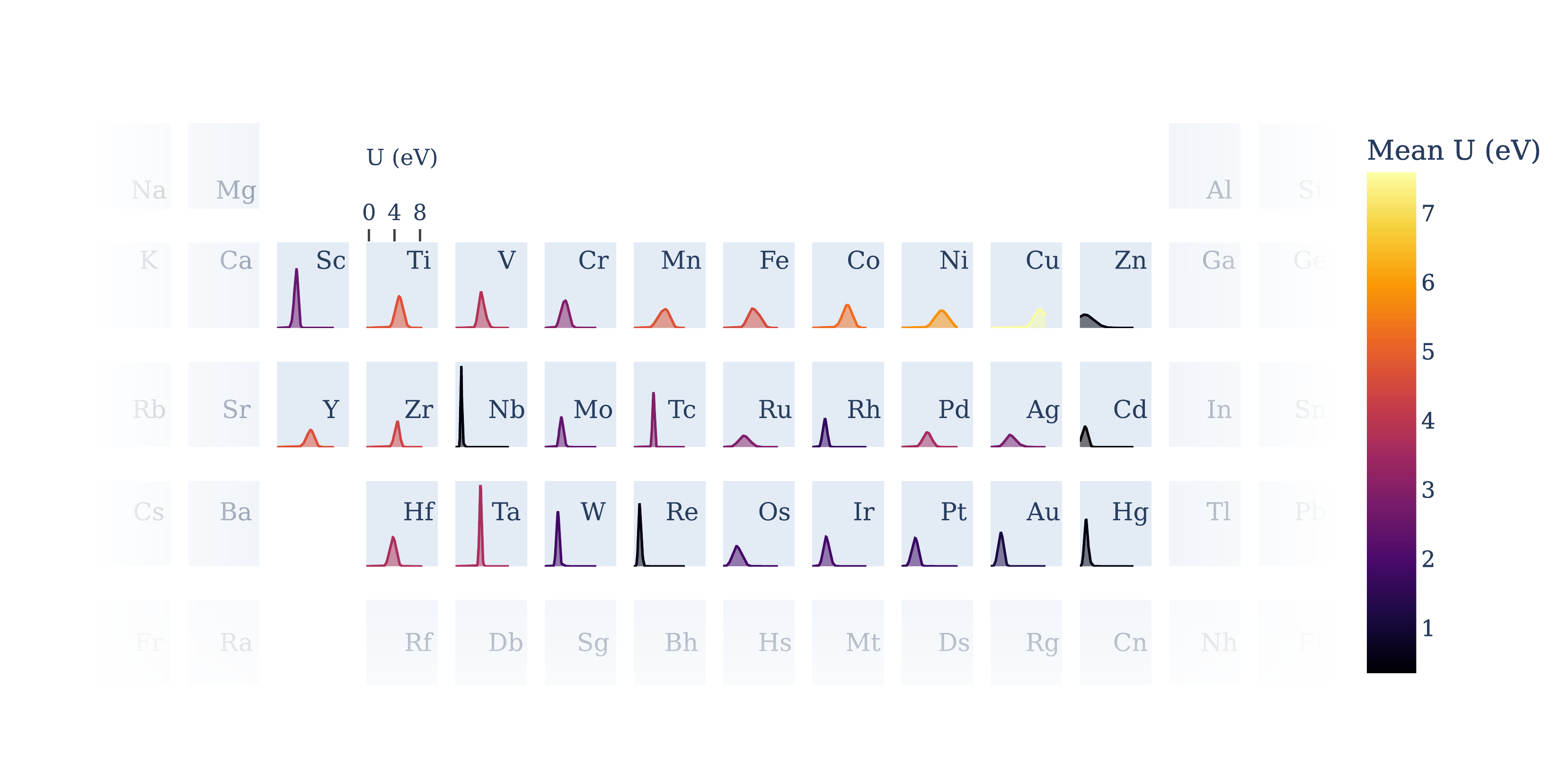}}} \\
    \subfloat[\centering Hund $J$ Periodic Table]{{
    \label{fig:ptable_j}
    \includegraphics[width=0.90\linewidth]{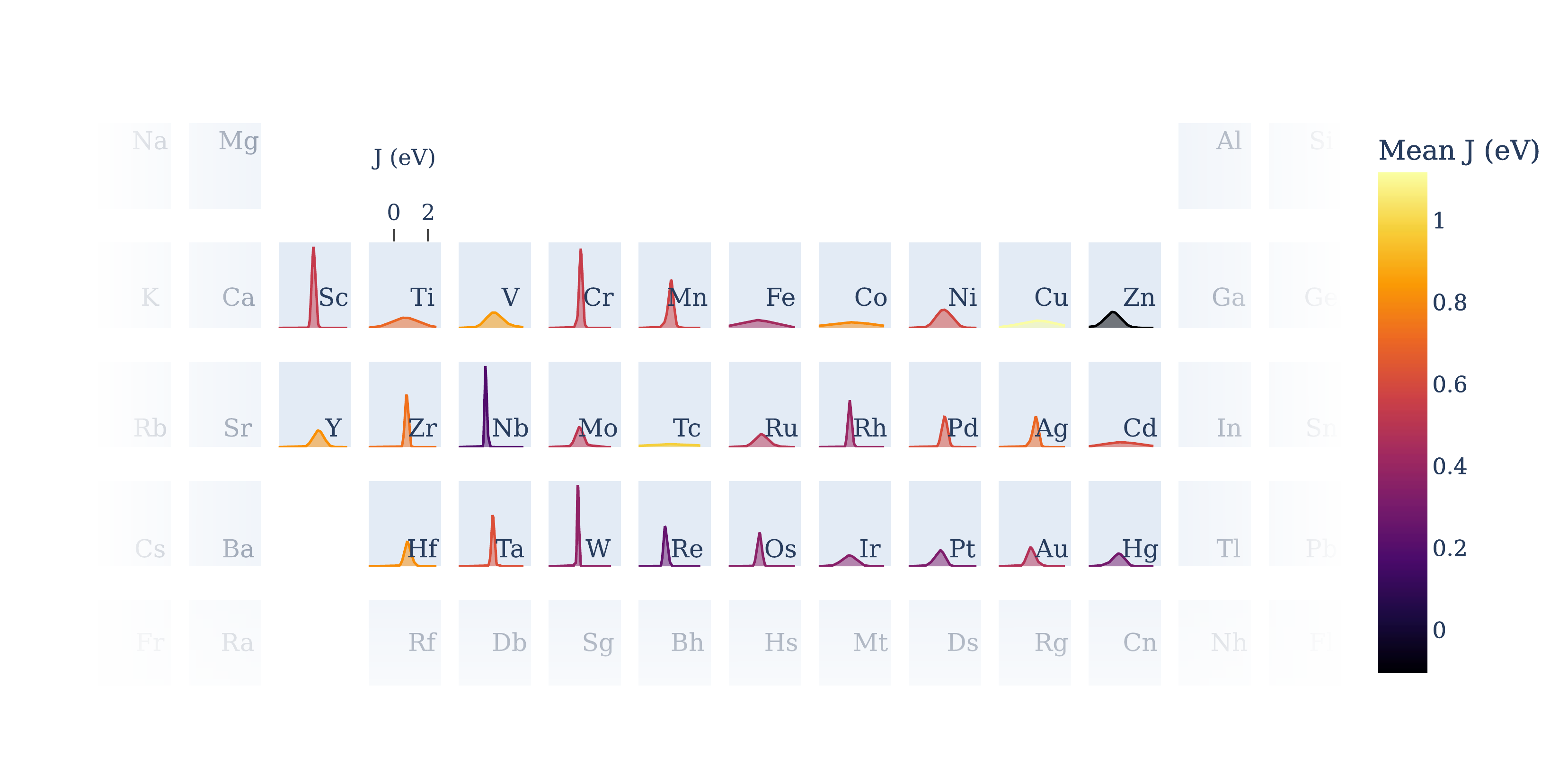}}}
    \caption{Periodic table of Hubbard $U$ and Hund $J$ values computed for representative set of transition metal oxides. The color map indicates the mean value computed for each element over each material. The materials used in the creation of these periodic tabled were selectively chosen: noting that many databases, including the ICSD, contain a growing number of hypothetical materials which may or may not be realizable, we selected materials that are well-studied and exhibit more than two ICSD IDs each. Furthermore, to remove cross-correlations between magnetic elements, we also require that these compounds only contain a single $d$-block element (occupying a single symmetrically-equivalent site) with no $f$-block species.
    Ultimately this data corresponds to the $U$ and $J$ values for over 800 materials, and are distributed over the transition metal species. A more detailed table containing data on the distribution of values is included in Appendix \ref{sec:sample_details}. The plotted distributions of $U$/$J$ values are generated using a Gaussian kernel-density estimator implemented in \texttt{scipy} \cite{scipy}.}
    \label{fig:ptable}
\end{figure*}

\begin{table}
\caption{Comparison of computed $U_{\text{eff}}$ in the present work with values used by the Materials Project~\cite{mp-uvals,cococcioni-ceder-ldau-redox}.}
\label{tab:mp_comparison_full}
\begin{ruledtabular}
\begin{tabular}{lccc}
element	&	mean $U_{\text{eff}}$ (eV)			& $U_{\text{eff}}^{\text{MP}}$ (eV)	&	diff. (eV)	\\% &	Consistent Within Error\\
\hline							
Co	&	4.430	$\pm$	1.474	&	3.32	&	1.110	\\% &	Yes\\
Cr	&	2.425	$\pm$	0.472	&	3.7		&	-1.275	\\% &	No\\
Fe	&	4.108	$\pm$	1.322	&	5.3		&	-1.192	\\% &	Yes\\
Mn	&	4.135	$\pm$	0.724	&	3.9		&	0.235	\\% &	Yes\\
Mo	&	1.911	$\pm$	0.318	&	4.38	&	-2.469	\\% &	No\\
Ni	&	5.258	$\pm$	0.773	&	6.2		&	-0.942	\\% &	No\\
V	&	3.060	$\pm$	0.673	&	3.25	&	-0.190	\\% &	Yes\\
W	&	1.461	$\pm$	0.218	&	6.2		&	-4.739	\\% &	No\\
\end{tabular}
\end{ruledtabular}
\end{table}

\subsection{Periodic table sample set}
\label{sec:periodic_table}

%\textcolor{red}{Is there anything we want to say about the periodic tables?}

Figure \ref{fig:ptable} displays two periodic tables containing the distributions of computed Hubbard $U$ and Hund's $J$ values for each transition metal element (and oxygen) computed for different structures within the database. In Table \ref{tab:mp_comparison_full}, values obtained in this study are listed alongside the standard $U$ values employed by the Materials Project \cite{mp-uvals,cococcioni-ceder-ldau-redox}. Those values were determined using the procedure outlined by Wang et al. \cite{oxidation-energy-GGAplusU} which finds a $U$ value that minimizes the error in formation energy for several representative redox couples. Due to the limited amount of experimental data available, these $U$ values are determined with only experimental data from a single redox couple (Co, Cr, Mo, Ni, and W) or two redox couples (Fe, Mn, and V). Therefore, it is possible or likely that these $U$ values are not appropriate for a more general system containing these elements. Nevertheless, the MP $U$ values are found to be the same as the $U$ values in the present work within the standard deviation for most elements (Co, Fe, Mn, and V) or slightly outside the value in the present work (Ni). Exceptions are Cr, Mo, and W, with the largest, notable discrepancy of 4.739 eV for W.

To evaluate the impact of these discrepancies, compounds containing W from a dataset of experimental formation energies \cite{Wang-Persson-2021} used by the Materials Project were taken and relaxed using the new $U_{\text{eff}}$ value for W from the present work but with all other calculation settings kept consistent with standard Materials Project settings, to obtain a new set of computed energies. These energies substantially lowered the correction introduced in Ref.~\onlinecite{Wang-Persson-2021} for W from -4.437 eV/atom to 0.12 eV/atom, suggesting that the newer $U_{\text{eff}}$ is indeed more appropriate for the calculation of formation energies.

\begin{table*}
\caption{Computed range of $U$, $J$, and $U_{\text{eff}}$ values compared with reported $U_{\text{eff}}$ on the Materials Project (MP) \cite{mp-uvals}, as well as the MP literature \cite{cococcioni-ceder-ldau-redox}. Each mean value has an associated standard deviation indicated after the ``$\pm$."}
\label{tab:mp_comparison}
\begin{ruledtabular}
\begin{tabular}{cccccc}
	&		mean		&		mean		&		mean		&	reported	&	reported	\\
	&	computed			&	computed			&	computed			&	MP \cite{mp-uvals}	&	range \cite{cococcioni-ceder-ldau-redox}	\\
Species	&	$U$ (eV)			&	$J$ (eV)			&	$U_{\text{eff}} = U - J$ (eV)			&	$U_{\text{eff}}$ (eV)	&	$U_{\text{eff}}$ (eV)	\\
\hline											
{Mn-$d$}	&	4.953	$\pm$	0.635	&	0.520	$\pm$	0.156	&	4.433	$\pm$	0.654	&	3.9	&	3.60 -- 5.09	\\
{Fe-$d$}	&	4.936	$\pm$	0.700	&	0.177	$\pm$	0.367	&	4.759	$\pm$	0.790	&	5.3	&	3.71 -- 4.90	\\
{Ni-$d$}	&	5.622	$\pm$	1.221	&	0.399	$\pm$	0.434	&	5.223	$\pm$	1.296	&	6.2	&	5.10 -- 6.93	\\
{O-$p$}	&	10.241	$\pm$	0.910	&	1.447	$\pm$	0.171	&	8.794	$\pm$	0.926	&	N/A	&	N/A	\\
\end{tabular}
\end{ruledtabular}
\end{table*}

We stress that these values are not transferable to other studies, which use DFT implementations in other codes. \textsc{Quantum ESPRESSO} and \textsc{Abinit} use localized projections that are different from the projector augmented wave (PAW) method implemented in VASP \cite{wang_local_2016}. 

%\textcolor{red}{Maybe say something about transferability/qualitative agreement?}

\subsection{Focused study on {Mn-$d$}, {Fe-$d$}, {Ni-$d$}, and {O-$p$}, including
the reason for large O-$p$ Hubbard $U$ values}
\label{sec:MnFeNiO}
\begin{figure*}
    \centering
    \subfloat[\centering Mn-$d$]{{
    \label{fig:Mn_UJ_m}
    \includegraphics[width=.32\linewidth]{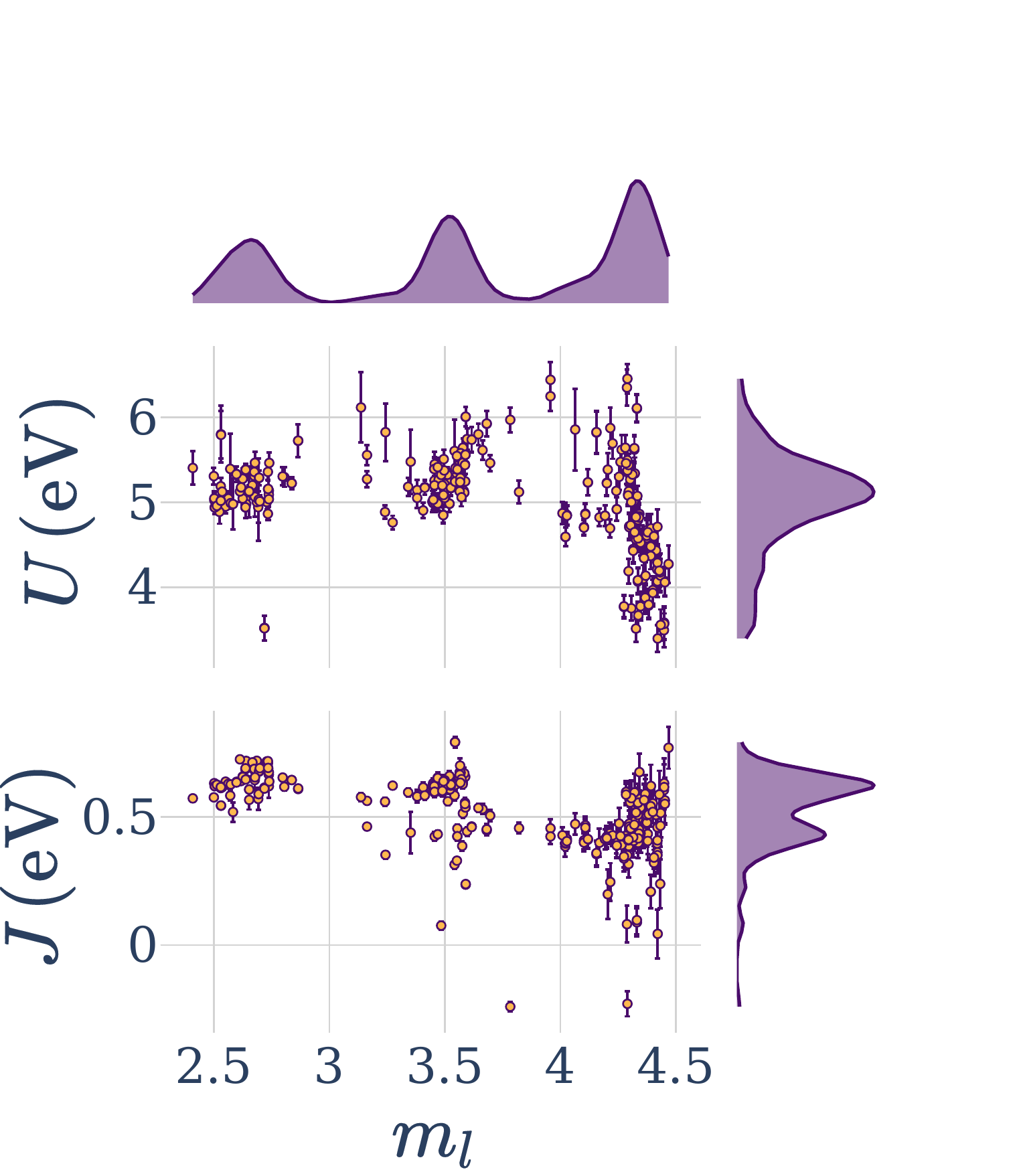} }}
    \subfloat[\centering Fe-$d$]{{
    \label{fig:Fe_UJ_m}
    \includegraphics[width=.32\linewidth]{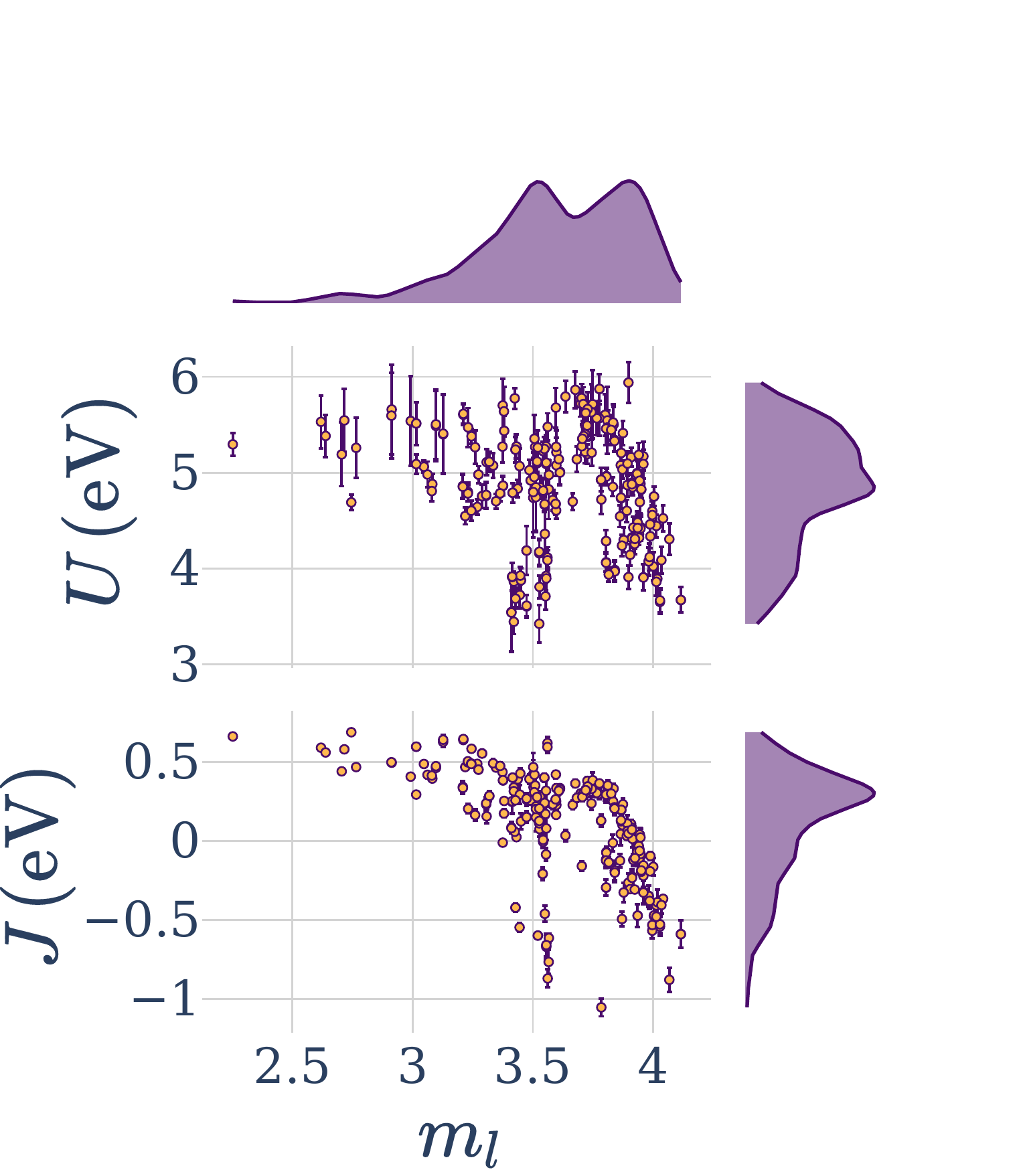} }} \\
    \subfloat[\centering Ni-$d$]{{
    \label{fig:Ni_UJ_m}
    \includegraphics[width=.32\linewidth]{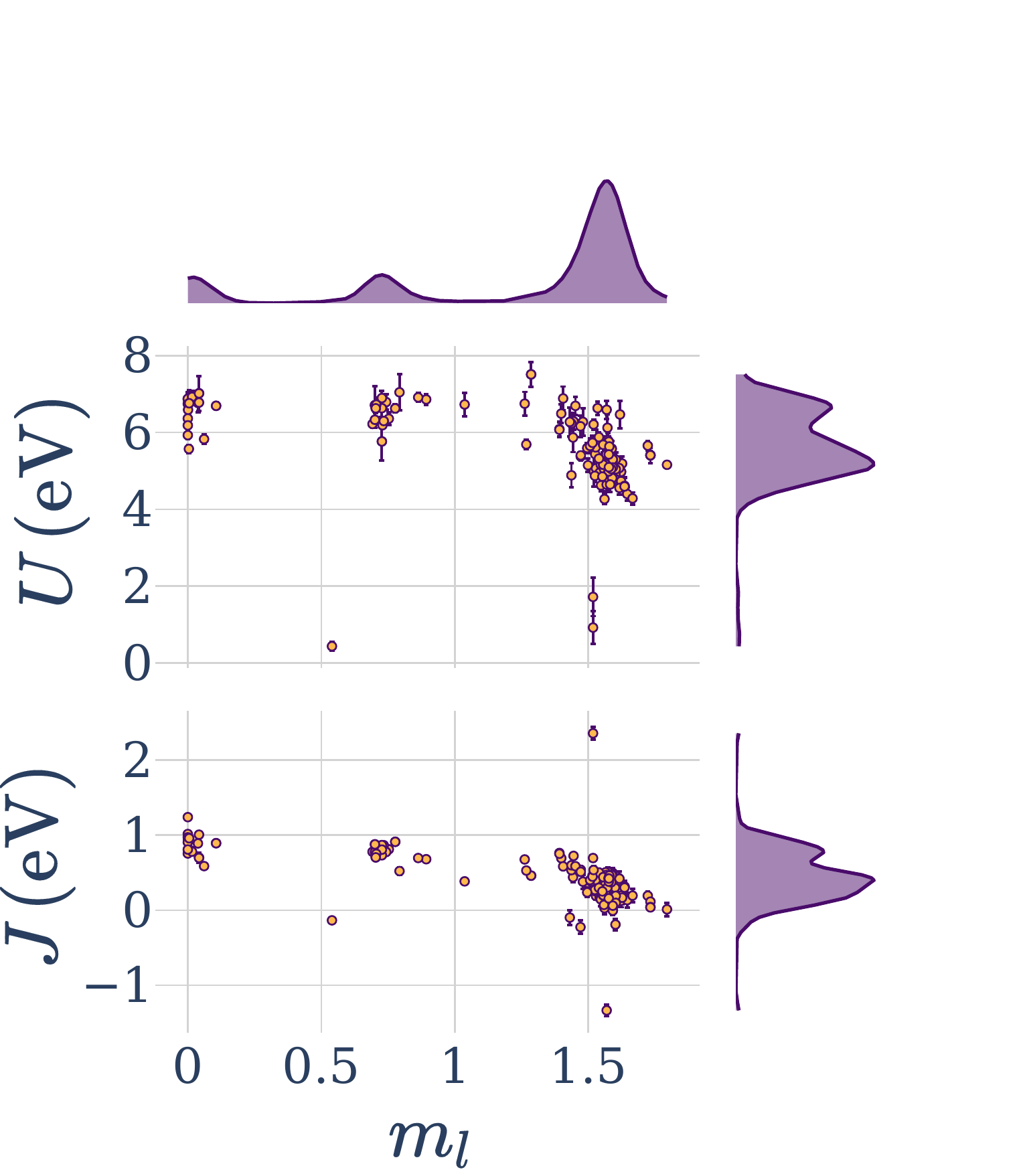} }}
    \subfloat[\centering O-$p$]{{
    \label{fig:O_UJ_n}
    \includegraphics[width=.32\linewidth]{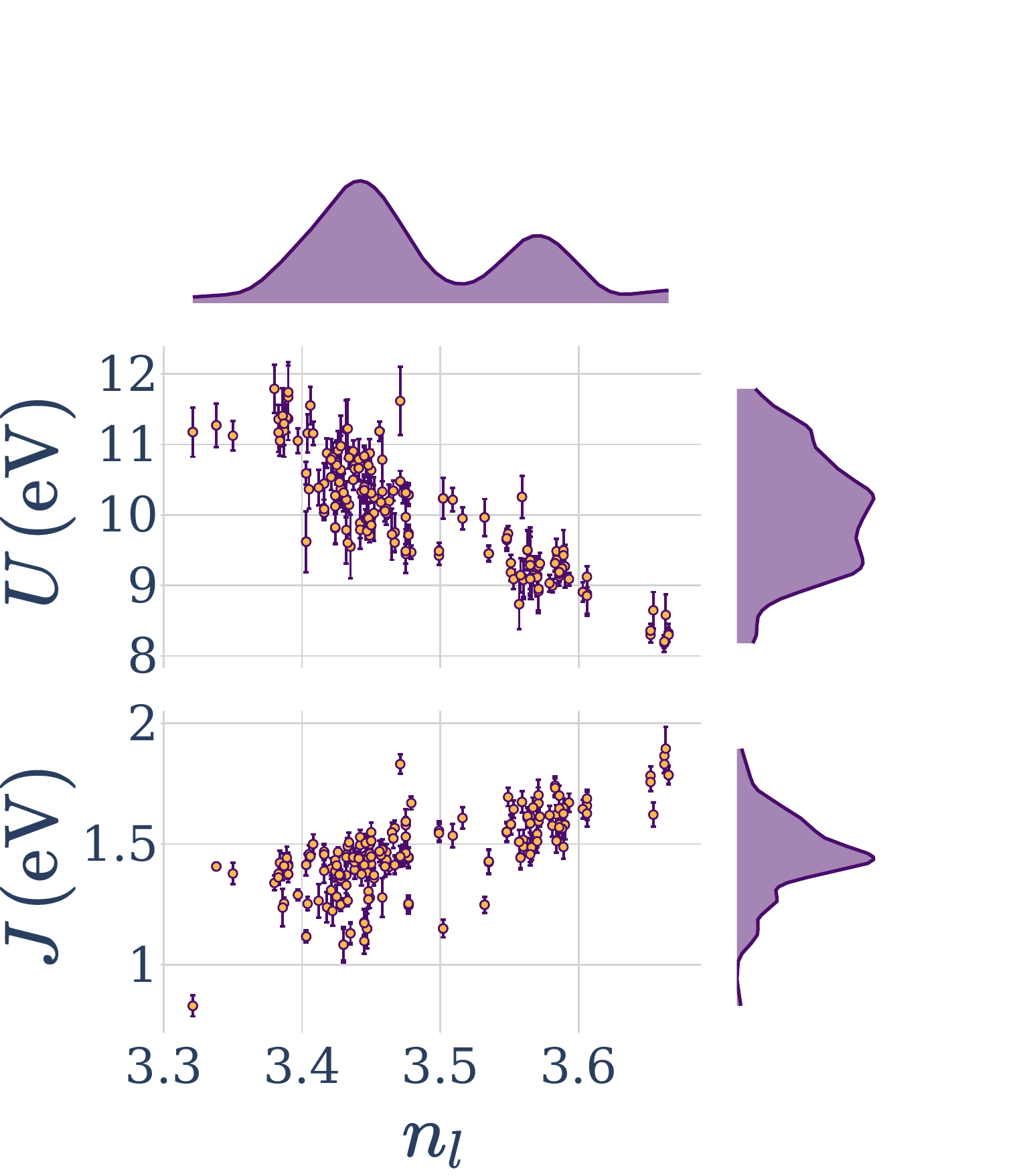} }}
    \caption{Distributions of Hubbard $U$ and Hund $J$ values computed using the linear response method; For the sub-figures (a), (b), and (c) that correspond to $d$-electron TM site corrections, the $U$ and $J$ values are plotted against the DFT (no +$U$+$J$ correction) computed site magnetic $m_{l}$, where $m_2$ and $m$-$d$ are equivalent because $d$ occupations have a corresponding $l=2$ angular momentum quantum number. The O-$p$ $U$ and $J$ values in sub-figure (d) are plotted against $n_l$ ($n_1$ or $n$-p) total site occupations. The number of samples for on-site correction values for Mn-$d$, Fe-$d$, Ni-$d$, and O-$p$ are 285, 248, 149, and 206, respectively.}
    \label{fig:UJ_vs_mn}
\end{figure*}

We now present a more detailed study on materials containing {Mn-$d$}, {Fe-$d$}, {Ni-$d$}, and {O-$p$} Hubbard sites. For these systems, the distributions of the computed Hubbard $U$ and Hund $J$ values are provided in Figure \ref{fig:UJ_vs_mn}. The variations in $U$ and $J$ values calculated for these three species is immediately apparent, with a range on the order of approximately 1 to 2 eV. These distributions reflect the intrinsic screening environment dependence of the calculated value for a 
given element.
At this point, we note only their 
apparently universal unimodality (single peak)
and the near-general decrease in $U$ with chemical
period within a given group, however
we will return presently to a more physically
and chemically motivated observation.
In Table~\ref{tab:mp_comparison} we list for comparison the $U$ values currently used in Materials project (fitted empirically) as well as a range of $U$ values found for a set of spinels and olivines by Zhou and co-workers (calculated via self-consistent linear response) \cite{cococcioni-ceder-ldau-redox}.
% In the work by Zhou et al. \cite{cococcioni-ceder-ldau-redox}, the effective Hubbard $U$ values were computed using the self-consistent linear response method developed by Cococcioni et al. \cite{corr-el-book-2012-ch4} for a collection of a few Mn, Fe, or Ni containing spinels and olivines. They present individual $U$ values for each expected charge state of Mn, Fe, or Ni. In addition, the $U_{\text{eff}}$ values reported for Materials Project are included below as well \cite{mp-uvals} in Table \ref{tab:mp_comparison}. These values roughly coincide with the ranges found by Zhou et al. \cite{cococcioni-ceder-ldau-redox}. $U_{\text{eff}}$ was computed as the difference between the mean of the computed $U$ and $J$ values.

We find that O-$p$ exhibits the largest associated Hubbard $U$ value of approximately 10 eV, which agrees with the linear response results from a previous study using a different code and 
somewhat different linear-response formalism \cite{LinscottEtAl}. 
While large oxygen Hubbard $U$ values may seem surprising within a 
strongly correlated materials context, it
has become more accepted in recent years within first-principles solid-state chemistry that 
oxygen 2p orbitals can warrant, both  by direct calculation and by necessity (when resorting
to fitting), a remarkably high $U$ value in
DFT+$U$.

We will now attempt to motivate and explain this
phenomenon.
We note from the outset that the  element projector orbital profile plays a complicating role in the following analysis.
In general, we observe that the diagonal elements of the $\chi_0$ non-interacting response matrix are of roughly the same magnitude for both TM-$d$ and O-$p$ sites. 
The non-self-consistent response can be interpreted as the response due to 
non-interacting response effects at a site due to its surroundings \cite{corr-el-book-2012-ch4}, and thus it can
be understood as a property primarily of the environment of the atom under scrutiny. 
Then, unless screening is very short ranged as it may be in a very wide-gap insulator, 
this quantity may be said to be somewhat similar, on average, for 
metal and oxygen ions in an oxide. 
Thereby, the chemical trends in the Hubbard $U$
arise mostly in the interacting response.
% and by deduction in their different, the interaction part.

Next, we note that the O-$p$ interacting response $\chi$ tends to be less than half of that of the interacting TM-$d$ response. This indicates that $-\chi^{-1} = d^2 E / d n_I^2$, the curvature of the total energy versus occupation, $n_I$, is greater for O-$p$ states. This greater curvature versus occupation can be explained, 
we propose, in terms of known 
trends in the chemical hardness, i.e., the second
chemical potential, i.e, 
the derivative of the chemical potential with
respect to total charge at fixed external potential.
We note, in passing, that some
authors choose define the chemical hardness
as half of that for historical reasons, but we suppress that here.
Specifically, we can focus on the discretized
(three-point) approximation 
to the global chemical hardness~\cite{doi:10.1021/ja00364a005}, 
namely 
\begin{align} \nonumber \nu &{}\equiv d^2 E / d N^2 \approx E\left(N-1\right) - 2 E\left( N \right) + E\left(N+1\right) \\ \nonumber &{}= 
\left[ E\left( N+1 \right) - E\left(N\right) \right] - \left[ E\left(N\right) -  E\left( N-1 \right) \right] \\
&{}\equiv E_i - E_a \equiv E_g,
\end{align}
which is nothing but the fundamental band-gap. 
This  is a  quantity 
that has been tabulated many times, 
and using the results of Ref.~\onlinecite{doi:10.1073/pnas.2117416119}
we find that for atomic oxygen its value is $11.2$~eV, compared to that of the transition metal atoms, 
where it ranges from $5.8$~eV (Ti \& Zr) to $8.0$~eV (Mn) if we exclude the
often problematic zinc group, where it reaches
$11.6$~eV.
%CITE
This mirrors and explains the observed relatively large first-principles Hubbard $U$ value for
oxygen 2p states predicted in this and
several previous studies.

% NO NEED TO CITE NOW
%\cite{allen-electronegativity-1989, mann-configuration-2000, mann-configuration-2000-1}. 
%Additionally, in a recent \DFTUJ{} study on crystal phases of \ce{TiO2} \cite{oregan-UJ-TiO2}, it has been shown that oxygen 2$p$ states are more localized than Ti 3$d$ states.
Ultimately, we conclude that
the Hubbard $U$ may be interpreted
as the subspace-projected, environment screened
chemical hardness, 
and more precisely as only the 
interaction (e.g., Hartree, exchange, 
correlation, and perhaps other terms
like implicit solvent and PAW potential) 
part of that. 
It is in the
interaction part that most of the
chemical trends appear to arise in practice.
For subspaces projecting heavily 
at both band-edges, as in normal
DFT+$U$ practice, the $U$
clearly inherits chemical
trends from the chemical hardness 
(fundamental gap) of
the atom that it resides upon. 
This is higher 
for a higher atomic ionization energy $E_i$ 
(that of oxygen is generally around twice that 
of transition metals) and higher also 
for a more negative 
electron affinity $E_a$ (that of oxygen is
more negative than that of most but not all
transition metals).
By and large, both quantities are well 
known to increase in magnitude
as we move `up and right' in the periodic
table, and this same broad trend is reflected
in our periodic table of Hubbard $U$ values.

When a DFT+$U$ subspace projects only onto
one or other band edge, as seems more 
commonly the case 
for charge-tranfer insulators, then then the
trend in only one of the ionization
energy and electron affinity will be very relevant
to the trends in $U$.
In the case of oxygen 2p orbitals projectors, 
due to the
electronegativity of oxygen typically there
will be little weight at the conduction band
edge, and so it is the (particularly clear)
trend in ionization energy that drives 
the relatively large $U$ value for oxygen.
Indeed, if this argument holds then one would guess
that the oxygen 2p $U$ value is roughly twice that
of an average transition-metal d subspace, which
turns out to be the case from first principles
linear response. 

The Hund's $J$, within the present formalism, 
may be interpreted as an analogue  
for the spin degree of freedom, and specifically
as minus (by a convention thought to originate with Ising) the interaction part of the 
subspace-projected, environment screened
spin-hardness, even the global atomic
version of which~\cite{GUERRA200637} has been a much less
thoroughly studied quantity.
The effect of choosing whether these subspace charge (spin) hardness quantities, the $U$ and $J$, are calculated in a fully relaxed manner, or with 
with a simulated fixed spin (charge), 
is explored in our comparison between 
conventional (scaled) and 
constrained (simple) spin-polarized linear 
response, respectively, below.

% \textcolor{red}{We need to introduce Fig 2 properly. This paragraph has been moved from later on and needs massaging into place} 

In order to explore trends in the distribution of $U$ and $J$ values, we have plotted these on-site corrections in scatter plots within Figure \ref{fig:UJ_vs_mn}. These plots illustrate the relationship between $U$ and $J$ values with respect to site occupations. For transition metal species, we plot $U$ and $J$ versus $m_{l=2}$, the ``$d$" component of the projected moment, $m$, denoted as ``$m_{l=2}$." These moment values are those output by VASP as the difference of up and down spin site occupancy numbers computed using PAW site-projection operators. Because the oxygen atoms do not have an associated magnetic moment, we plot O-$p$ Hubbard $U$ and Hund $J$ versus $n_{l=1}$ occupations on oxygen sites. 

We should stress that the values of ``$m_{l}$" and ``$n_{l}$" are only computed from the calculation without the $+U$ correction. One reason for using the bare PBE computed $m_{l}$ and $n_{l}$ is that these occupations should be independent from the applied Hubbard $U$ or Hund $J$ values. This would offer the ``bare" $m$, as well as $n$, as a possible predictors of $U$ and $J$ values. However, it is important to note that these occupations could change significantly with applied $U$ and $J$ values \cite{LinscottEtAl, scf-u-SrMnO3, doi:10.1063/1.4865831}.

There is an apparent clustering of data points at different on-site $m_{l}$ magnetizations in Figures \ref{fig:Mn_UJ_m}, \ref{fig:Fe_UJ_m}, and \ref{fig:Ni_UJ_m}. This grouping at different on-site magnetization values is most likely due to different spin and charge states dependent on the underlying chemistry. We also observe a larger range of $U$ and $J$ values for higher values of $m_{l}$, which is due to the coupling between highly spin-polarized states to on-site Coulomb screening for TM species. As would be expected, we see similar trends for $J$, a measure of the screened interaction between spin channels.

For the Mn-$d$ and Ni-$d$ distributions in Figures \ref{fig:Mn_UJ_m} and \ref{fig:Ni_UJ_m}, a stark clustering of data-points is evident at particular intervals of $m_{l=2}$. In both cases, the clusters that lie at the associated maximum computed $m_{l}$ fall off and exhibit a negative slope trend with the magnitude of the site moment. This is likely due to the fact that $m_{l}$ is highly dependent on the local chemical environment, which will govern the energy curvature over spin occupations, which is directly related to $U$ and $J$ within linear response \cite{PhysRevB.71.035105}. The clear trend for the manganese may be due to the strong tri-modal distribution of Mn magnetic moments seen in Figure 1 of Ref.~\onlinecite{horton-high-throughput-2019}. The ``stable" magnetic configurations from this study were used in the LR analysis, therefore a similar statistical distribution should hold for the subset of structures used in this LR analysis.

The trends of the data points for Hubbard $U$ and Hund $J$ values in Figure \ref{fig:O_UJ_n} appear to show a downward trend for $U$ versus $p$-occupation numbers, $n_{l=1}$, and a slower, upward trend for $J$ values versus $n_{l=1}$.  We expect that the $n_{l=1}$ occupations will be strongly dependent on the oxidation/reduction state of oxygen atoms. Due to the nature of TM-O bonding in these oxides, and
their generally greater electronegativity, the oxygen atoms will tend to maximize their valence. Therefore, building on the previous explanation of the magnitude of O-$p$ $U$ values based on chemical hardness and  specifically the more relevant 
ionization potential component of that, the higher electron count for oxygen corresponds to a lower ionization potential, and therefore to a reduced  Hubbard $U$, as observed.

In an attempt to more robustly tease apart these observed trends, we performed a rudimentary random forest regression test on the data set, ultimately in an attempt to predict the on-site corrections $U$ and $J$ from the input crystal structures and site properties. We used the random forest regression algorithm as implemented in \texttt{scikit-learn}. The input quantities supplied to the random forest regressor consisted of the corresponding PBE-computed $m_{l}$ and $n_{l}$ - without on-site corrections, as well as the oxidation state estimated using the bond-valence method \cite{bvm-method-paper}, and finally a selection of relevant site featurizers provided by the \texttt{matminer} Python package \cite{matminer-paper}. Unsurprisingly the $U$ and $J$ values appeared to be the most sensitive to the magnetic moment magnitude, $m = n_\uparrow - n_\downarrow$, and site occupation, $n = n_\uparrow + n_\downarrow$. This is in accordance with what would be expected from the dependence on the Hubbard $U$ values on spin and charge state \cite{scf-u-SrMnO3, doi:10.1063/1.4865831}. However, these features proved to be insufficient to accurately predict $U$ and $J$.

Most of the \texttt{matminer} site featurizers were tested as input to the random forest regression model. Ewald energy and Voronoi site featurizers had the greatest associated importance metric \cite{matminer-paper}, second to $m_l$. However, the associated importance values of these featurizers were still less then the on-site magnetization, $m_l$. Additionally, the oxidation states calculated using the bond valence method (BVM) \cite{bvm-method-paper} were also included as input to the model. These guessed oxidation states are also used as input for the Ewald site featurizer. For learning trends across different atomic species, the atomic number of the associated element was also supplied. Additionally, we tested the orbital field matrix (OFM) features as formulated by \cite{lamphamMachineLearningReveals2017, karamadOrbitalGraphConvolutional2020}. The OFM encodes the orbital character of the surrounding chemical environment. For more information on this method please refer to Ref.~\onlinecite{lamphamMachineLearningReveals2017}. The OFM functionality is not implemented in \texttt{matminer} or \texttt{pymatgen}. We were motivated to test the vectorized OFM by the chemical intuition that on-site Hubbard $U$ and Hund $J$ values are very sensitive to the local chemical environment. Additionally, the OFM has demonstrated success in predicting DFT-computed magnetic moments in the past \cite{lamphamMachineLearningReveals2017}. Furthermore, the OFM nearest-neighbor contributions are weighted according to the geometry of the Voronoi cell, which could possibly provide information beyond the relative importance of the Voronoi \texttt{matminer} featurizer. However, the on-site magnetization for Mn, Fe, and Ni, respectively, had an importance of at least ten percent more than any of the other local chemical environment descriptors.

The correlation between on-site corrections and projected site moments is not surprising. After all, previous studies have explored the connection between charge states of transition metal species and the integrated net spin calculated from DFT \cite{yang-approaches-2022,reed-role-2004,kang-synthesis-2003}. The integrated atomic spin moment can be directly linked to the charge state of transition metal species via magnetochemistry rules. In fact, recent studies show that the magnetic moment is often the most convenient and reliable indicator of charge states \cite{yang-approaches-2022}.

% \textcolor{red}{This section needs some % clearer and/or more compelling conclusions
% \begin{itemize}
%    \item There are some obvious trends of $U$ and $J$ vs $n$ and $m$
%     \item $U$ and $J$ are too specific to the local chemical environment to be determined by $n$, $m$ alone
% \end{itemize}
% }

\subsubsection{Conventional vs. constrained linear response}
\begin{figure}
    \centering
    \includegraphics[width=0.90\linewidth]{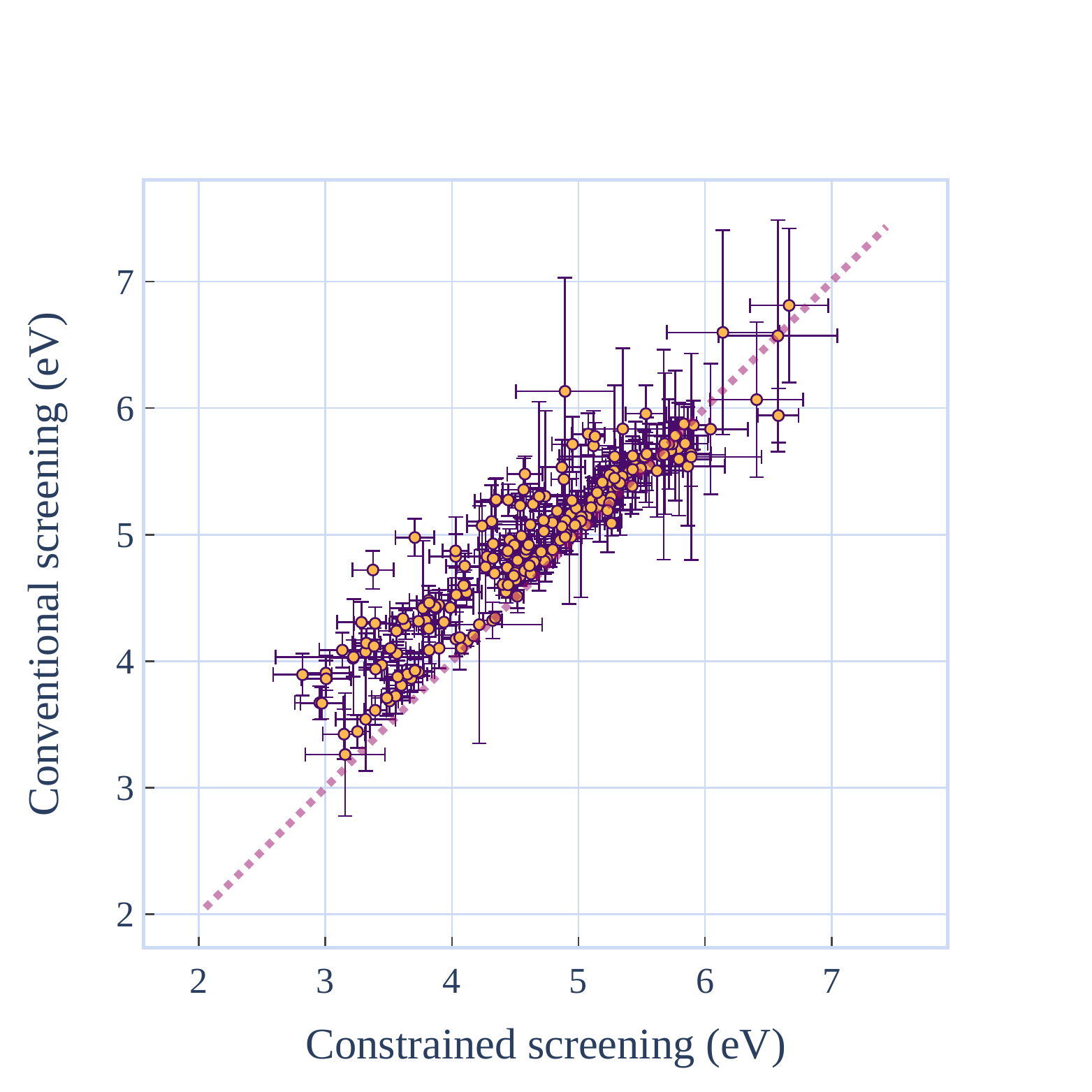}
    \caption{Comparison between the conventional and constrained approaches for calculating Hubbard $U$ values for {Fe-$d$} Hubbard sites.
    % (b) Comparison between ``atom-wise" versus ``full" screening inversion methods for Hund $J$ values calculated for {O-$p$}. These $J$ values are computed via the ``scaled" formula.
    }
    \label{fig:scalecompare_Fe_U}
\end{figure}

In introducing the linear response theory in Section \ref{sec:lr_theory}, we mentioned that there are two possible schemes for computing $U$ and $J$: ``conventional" and ``constrained" linear response, where in the latter case the linear response is performed in such a way that the magnetic moment (occupation) is held fixed while measuring the curvature with respect to the occupation (magnetic moment). While arguments can be made as to theoretically which approach is the most valid (a topic which is the subject of ongoing research), this dataset presents an opportunity to evaluate how much this choice will practically affect the resulting Hubbard and Hund's parameters.

For the majority of the computed $U$ and $J$ values using these two methods, the difference between the two strategies fell within their computed uncertainty. However, we observed a significant deviation from $y=x$ behavior for the computed $U$ values for iron Hubbard $U$ values shown in Figure \ref{fig:scalecompare_Fe_U}. The width of this distribution is greater than 1 eV for $U$ in some regions, which is enough to affect computed physical properties \cite{PhysRevB.71.035105, scf-u-SrMnO3}. 

\subsubsection{Dependence on structure and magnetic state}

For some input magnetic structures, the magnetic configuration changed while applying the on-site potentials during the linear response analysis. Our hypothesis is that the input magnetic structure corresponds to a local minimum configuration, or possibly a metastable state. Therefore, in our analysis, we screen out these structures with the intent that these systems will be studied in the future using a self-consistent approach to calculating on-site corrections.

In order to test the sensitivity of $U$ and $J$ values to the input structure, we perform a self-consistent linear response study of antiferromagnetic NiO, which is provided in the Supplementary Information. Each iteration consists geometry optimization of cell shape, followed by a linear response calculation of $U$ and $J$ values. These on-site correction values are then used in the next subsequent geometry optimization step. Self-consistency is achieved once the $U$ and $J$ values fall within their corresponding uncertainty values. Starting from the input structure --- which was optimized using the current default Materials Project $U$ values \cite{mp-uvals} --- convergence was achieved after only two iterations.

% \textcolor{red}{In closing, shall we make a general comment about importance of self-consistency and the possibility of an extension to this work? e.g. is it straightforward to incorporate self-consistency within the workflow}

It has been well established in previous studies that $U$ values should be computed self-consistently with geometry optimization \cite{scf-u-SrMnO3}. As demonstrated from the experiments with antiferromagnetic NiO in the Supplementary Information section, the Hund $J$ values should be calculated self-consistently, in addition to Hubbard $U$ values. In this self-consistency study, $J$ had the largest associated change over convergence relative to the value itself. Due to the coupling between Hund $J$ and magnetic exchange \cite{streltsov-orbital-phys-2017}, it is possible that both magnetic and structural features should be included in the self-consistency cycle. Within the \texttt{atomate} framework, it would be possible to incorporate a workflow that wraps the workflow developed in this study, in order to alternate linear response calculations with geometry relaxation until self-consistency is achieved.

\subsection{Case study: \ce{LiNiPO4}}
\label{sec:LiNiPO4}

\begin{figure}
    \centering
    \includegraphics[width=0.98\linewidth]{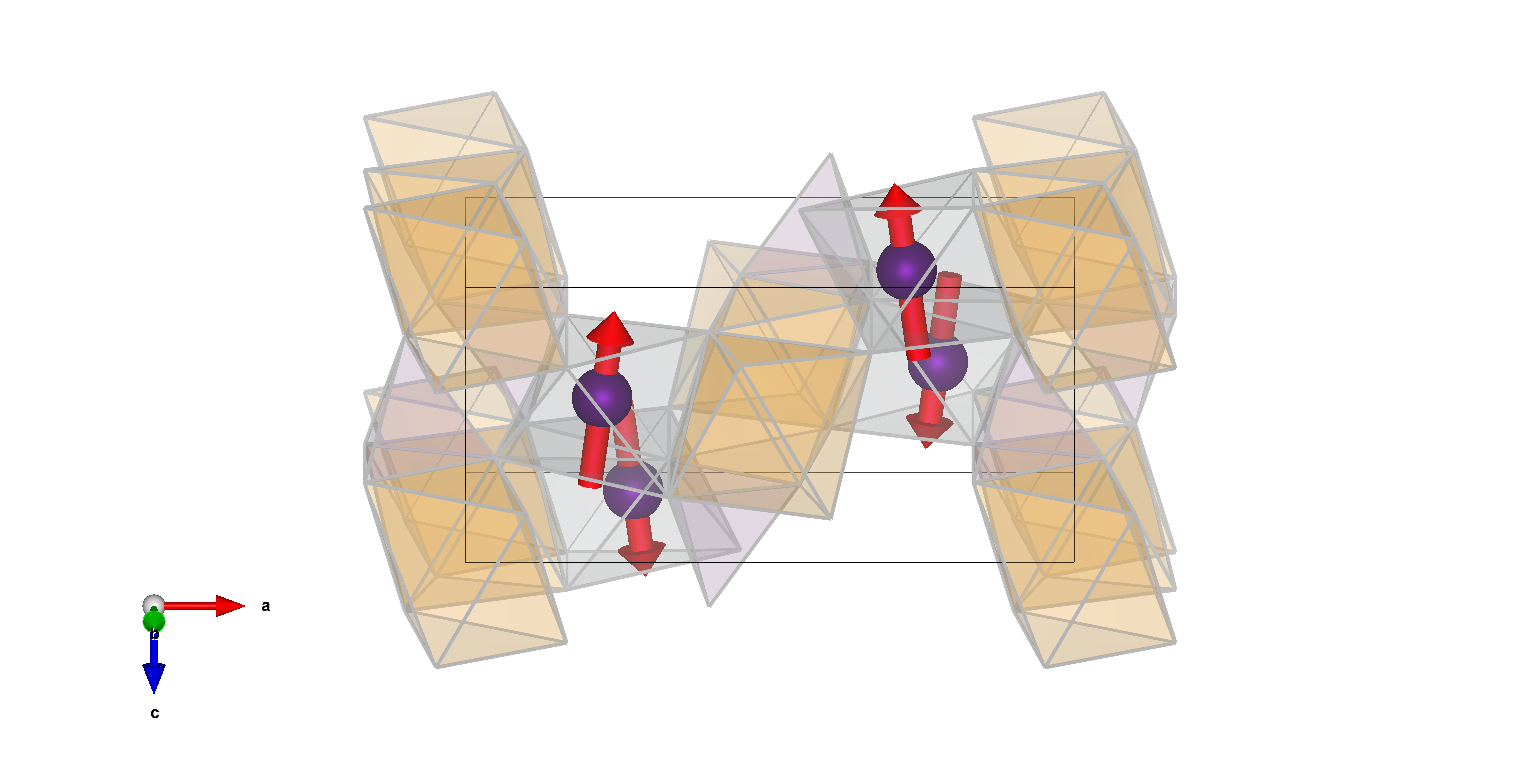}
    \caption{Olivine crystal structure of \ce{LiNiPO4} with magnetic atoms visible. Taken from \cite{exp-LiNiPO4} via the Bilbao MAGNDATA database \cite{bilbao-server, exp-LiNiPO4}. The purple atoms correspond to magnetic nickel atoms. The oxygen octahedra surrounding lithium atoms are indicated in orange, where the grey oxygen octahedra surround nickel sites.}
    \label{fig:exp1_struct_LiNiPO4}
\end{figure}

We now present a detailed study on the olivine \ce{LiNPO4}, designed to test the results produced by the linear response workflow. Previous GGA$+U$ and GGA+$U$+$J$ studies have attempted to reproduce the experimentally-observed spin-canting structure and unit cell shape as shown in Figure~\ref{fig:exp1_struct_LiNiPO4} \cite{GGAplusUJ-LiNiPO4,cococcioni-ceder-ldau-redox,exp-LiNiPO4}.

We calculated $U$ and $J$ for this system via spin-polarized linear response. The spin-polarized linear response method introduced in Section~\ref{sec:lr_theory} can be generalized to noncollinear DFT using the relationship between spin-density occupations and the magnitude of the magnetic moment: $n_\uparrow = \frac{1}{2}\left( n + |\vec{m}| \right)$ and $n_\uparrow = \frac{1}{2}\left( n - |\vec{m}| \right)$ \cite{dudarev-ncl-UO2-2019}. For comparison, we also performed a collinear calculation, where the magnetic configuration for \ce{LiNiPO4} was obtained by projecting the canted noncollinear structure shown in Figure \ref{fig:exp1_struct_LiNiPO4} along the $z$-direction. In addition to one unit cell of the the collinear antiferromangetic (AFM) configuration, a linear response analysis was performed on a 1$\times$2$\times$2 supercell. 
%The reasoning for computing the $U$ and $J$ values for a larger supercell can be justified by the results by Cococcioni and others \cite{PhysRevB.71.035105} that shows the strong influence of the computed $U_{\text{eff}}$ on the supercell size.
Table \ref{tab:computed_uj_Ni-d_LiNiPO4} summarizes the results of the computed Hubbard $U$ and Hund $J$ values. From this table, it is evident that the $U$ value is significantly smaller in magnitude with the inclusion of spin-orbit coupling. A possible justification for this behavior is the introduction of orbital contributions to the total localized magnetic moments with the inclusion of spin-orbit coupling \cite{PhysRevB.99.104421, streltsov-orbital-phys-2017}. 

% \textcolor{red}{What are our comments on this table, if any?}

\begin{table}
\caption{ Hubbard and Hund results for {Ni-$d$} in LiNiPO$_4$ (Atom-wise screening) }
\label{tab:computed_uj_Ni-d_LiNiPO4}
\begin{ruledtabular}
\begin{tabular}{ llcc }
cell  & magnetism & $U$ (eV)
  % & $U$ (eV)
  & $J$ (eV)
  % & $J$ (eV)
  \\ 
  % & \textit{scaled}
  % & \textit{simple}
  % & \textit{scaled}
  % & \textit{simple}
  % \\
  \hline
  $1\times 1\times 1$ & collinear
  & 5.43 $\pm$ 0.16 %5.42569178019408 %0.156414626478405
  % & 5.20 $\pm$ 0.15 %5.19932922465518 %0.146195616140657
  & 0.38 $\pm$ 0.07 %0.380558603284867 %0.067981104647761
  % & 0.496 $\pm$ 0.146 %0.495993986982553 %0.146195616140657
  \\
  $1\times 2 \times 2$ & collinear
  & 5.44 $\pm$ 0.24 %5.44469348274811 %0.241038086990673 
  % & 5.03 $\pm$ 0.19 %5.02659838787418 %0.185644969210575
  & 0.54 $\pm$ 0.07 %0.537163425644415 %0.0691336874234945
  % & 0.687 $\pm$ 0.186 %0.686858456450301 %0.185644969210575
  \\
  $1\times 1 \times 1$ & non-collinear
  & 5.09 $\pm$ 0.15 %5.09435143821926 %0.147327986137423
  % & 4.82 $\pm$ 0.13 %4.8180669273979 %0.126859643086191
  & 0.42 $\pm$ 0.05 %0.421883849487388 %0.0492892653114332
  % & 0.595 $\pm$ 0.127 %0.595160112091501 %0.126859643086191
  \\
\end{tabular}
\end{ruledtabular}
\end{table}

% We do not observe an appreciable difference between the $2 \times 2$ Hubbard $U$ values computed for the single cell and supercell. This suggests improved supercell scaling of VASP's implementation of localized projections and/or the type of pseudopotentials used in this study. The authors hypothesize that the poor scaling of linear response $U$ values found in previous studies \cite{PhysRevB.71.035105, bennett_systematic_2019} can be attributed to the use of ultrasoft pseudopotentials (USPPs). However, we are unable to test this hypothesis because USPPs cannot be used with on-site Hubbard corrections in VASP. The improved scaling behavior for the atom-wise inversion does not extend to the full matrix inversion, as shown in the experiment for \ce{NiO} in the Supplementary Information. The reason for the lack of scalability of the full inversion results will be an interesting subject for future studies to explore.

\subsubsection{Canting angle exploration}

\begin{figure*}
    \centering
    \subfloat[\centering Computed energy versus constraining canting angle]{{
    \label{fig:canting_energy_plot}
    \includegraphics[width=0.45\linewidth]{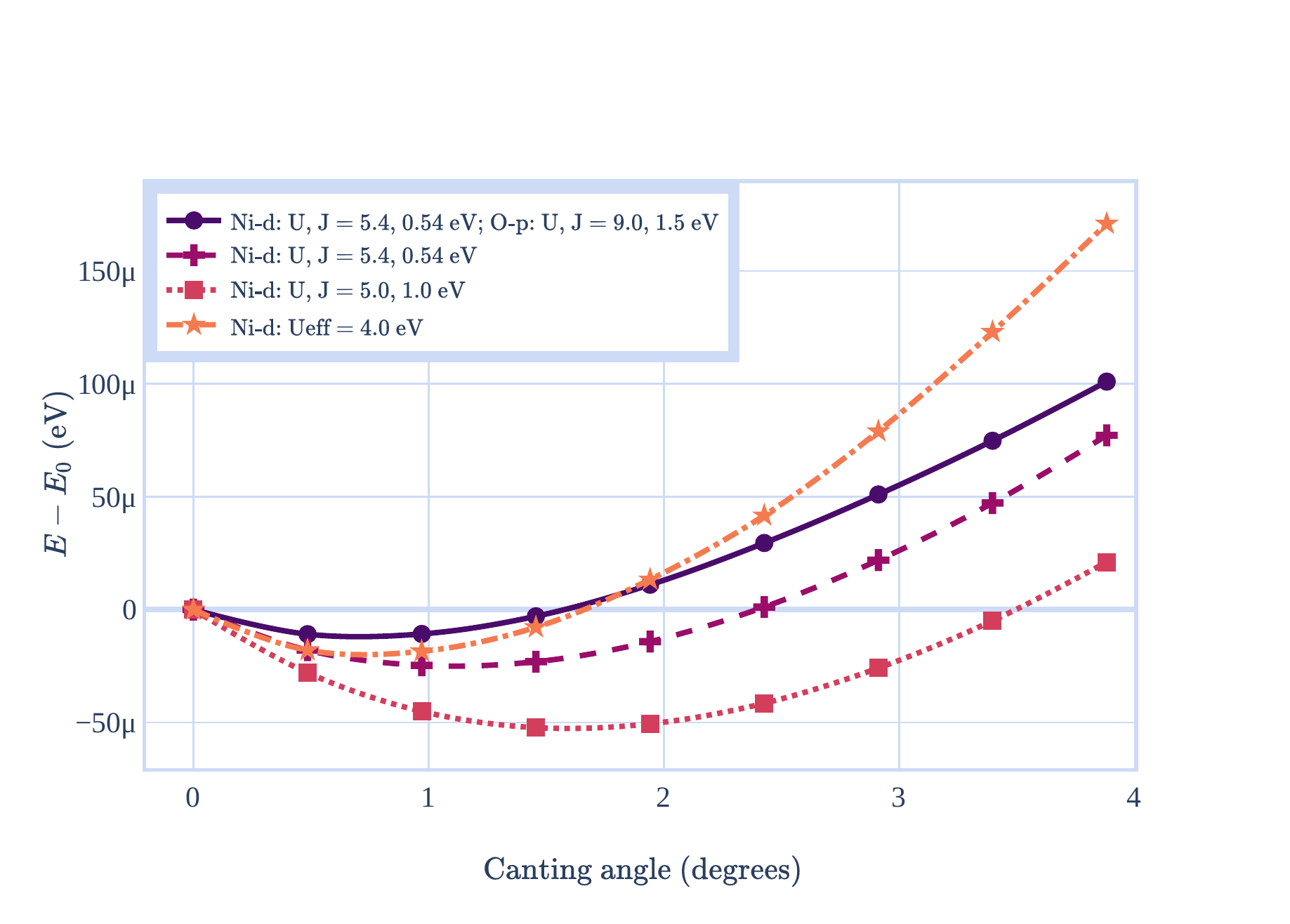}}}
    \subfloat[\centering Effective constraining field versus constraining canting angle]{{
    \label{fig:canting_Heff_plot}
    \includegraphics[width=0.45\linewidth]{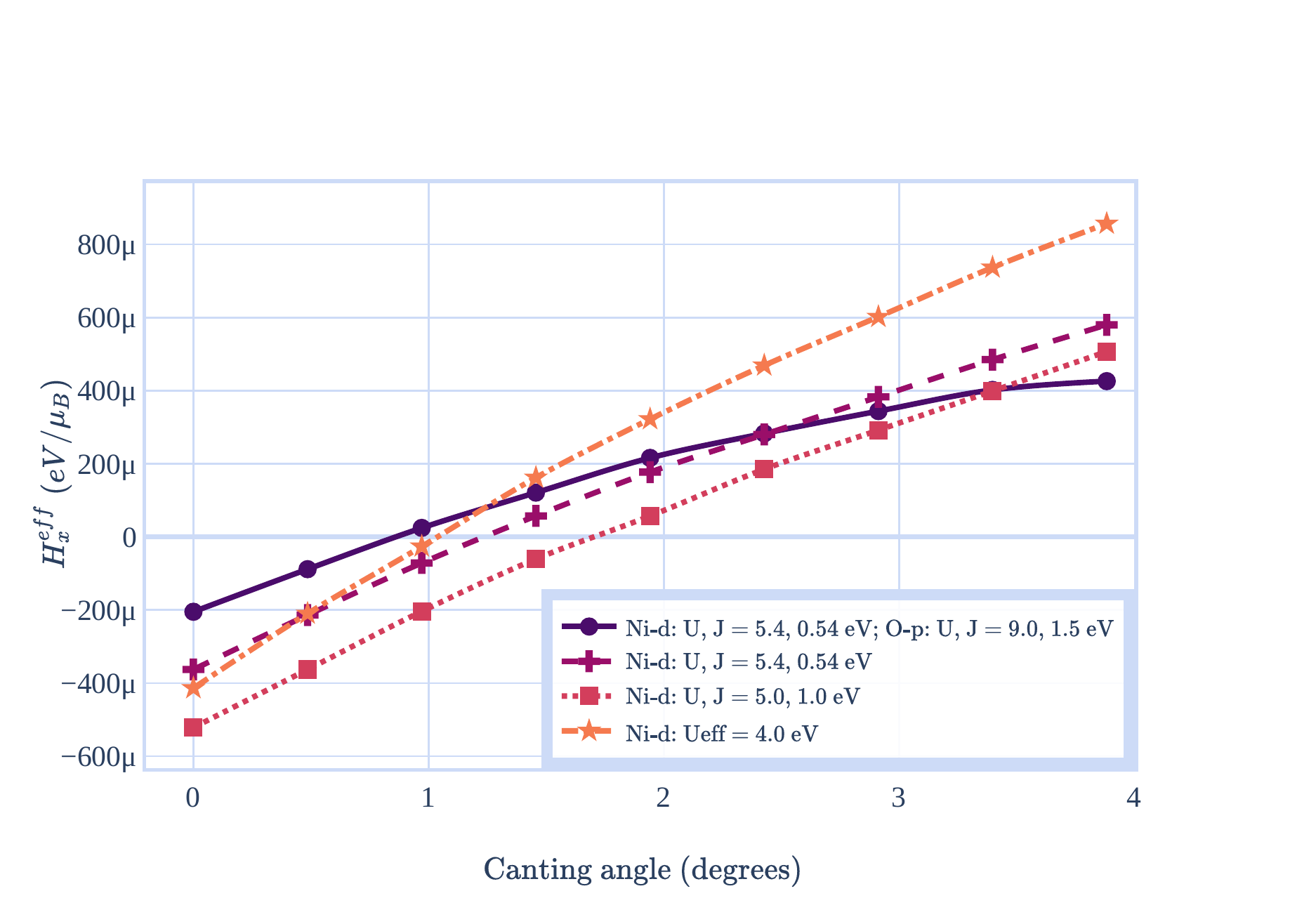}}}
    \caption{(a) Computed relative energy and (b) $x$-component of effective constraining local magnetic field for various Hubbard and Hund on-site corrections applied to the Ni-$d$ and O-$p$ manifolds.}
\end{figure*}

In order to explore the canting angle experimentally observed for \ce{LiNiPO4} \cite{exp-LiNiPO4}, we calculated the energy as a function of constrained canting angle. The noncollinear magnetic constraints were performed in VASP in accordance with the method developed by Ma and Dudarev \cite{constrain-ncl-vasp}. We used the experimentally derived spin canted structure as a reference provided by the Bilbao Crystallographic Server, as shown in Figure \ref{fig:exp1_struct_LiNiPO4} \cite{bilbao-server, exp-LiNiPO4}. The energy versus canting angle curve is shown in Figure \ref{fig:canting_energy_plot}. We found that the stable canting direction is in the opposite direction to the experimentally measured canting angle. However, this discrepancy with experiment was limited to the canting direction; the computed stable magnetic structure still obeyed the symmetry of the Pnm'a magnetic space group.

Similarly to the work by Bousquet and Spaldin \cite{GGAplusUJ-LiNiPO4}, we observe an increasing canting angle with Hund $J$ value. Interestingly, adding a $U$ and $J$ correction to O-$p$ results in a slightly decreased stable canting angle. However, we find that in all cases, the computed stable canting angle is significantly less than the experimentally measured canting angle of 7.8 degrees \cite{exp-LiNiPO4}.

The constraining effective site magnetic field, $\vec{H}^{\text{eff}}_i$, can be described as the following
\begin{align}
    \vec{H}^{\text{eff}}_i &= 
    2 \lambda \left[{\vec{M}}_i -{\hat  {M}}_i^0\left({\hat {M}}_i^0\cdot {\vec {M_i}}\right)\right]
\end{align}
where ${\vec{M}}_i$ are the integrated magnetic moments at site $i$, and ${\hat {M}}_i^0$ are the unit vectors pointing in the individual site constraining directions \cite{constrain-ncl-vasp}. The $x$ component of the constraining field (in the direction of canting), ${H}^{\text{eff}}_{i,x}$, is plotted versus the constraining angle in Figure \ref{fig:canting_Heff_plot}. We see that where ${H}^{\text{eff}}_{i,x}$ changes sign corresponds to the minimum of Figure \ref{fig:canting_energy_plot}.

\subsubsection{Effect of $U$ and $J$ values on geometry optimization}

\begin{table*}
\caption{ 
Lattice parameters, cell volume, and mean Ni-O bond length ($d$) of LiNiPO$_4$ canted structure for different Hubbard $U$ and Hund $J$ corrections
}
\label{tab:cellsize_UJ_LiNiPO4}
%\begin{ruledtabular}
\renewcommand*{\arraystretch}{1.5}
\footnotesize
\begin{tabular}{ l  l l l l l l l l}
\hline \hline
 method &  Ni-$d$ (eV) & O-$p$ (eV)
                       & $a$ (\angstrom) & $b$ (\angstrom) & $c$ (\angstrom) & volume (\angstrom$^3$) & $d$ (\angstrom) \\
\hline
\hline
experiment & 
&                       &    10.03      &    5.85       &    4.68       &    274.93    
&  2.086 $\pm$ 0.044 \\
\hline
  {PBE}
& 
&
                       &
    10.09  (+0.6\%)      &
    5.92   (+1.1\%)      &
    4.72   (+0.9\%)      &
    282.09 (+2.6\%)    &
     2.099 $\pm$ 0.037 \\
\hline
  \multirow{ 2}{*}{PBE$+U_{\text{eff}}$ } 
  & \multirow{2}{*}{$U_{\text{eff}}$ = 4} & $U_{\text{eff}}$ = 0 &
   10.14 (+1.1\%)      &
    5.92 (+1.1\%)      &
    4.73 (+1.0\%)      &
  283.71 (+3.2\%)    \\
%\hhline{~|--|----}
\cline{3-8}
& & $U_{\text{eff}}$ = 7.5 &
10.07   (+0.4\%) & 
 5.87   (+0.3\%) & 
 4.69   (+0.3\%) & 
 277.56 (+1.0\%) \\
\hline
  \multirow{ 2}{*}{PBE+$U$+$J$} & $U$ = 5
& $U$, $J$ = 0 &
     10.15 (+1.2\%)      &
      5.92 (+1.1\%)      &
      4.73 (+1.0\%)      &
    284.19 (+3.4\%)    &
  2.108 $\pm$ 0.039 \\
\cline{3-8}
& $J$ = 1 & $U$, $J$ = 9, 1.5 &
   10.07 (+0.4\%) &
    5.88 (+0.4\%) &
    4.69 (+0.3\%) &
  277.86 (+1.1\%) &
  2.095 $\pm$ 0.043 \\
\hline \hline
\end{tabular}
%\end{ruledtabular}
\end{table*}

% Note: scaling of energy for O-p U&J runs

% \begin{table*}
% \caption{Mean next-neighbor experimental Ni-O bond length, compared with computed relaxed structures with: no Hubbard or Hund corrections applied, as well as $+U$ and $+J$ applied to Ni-$d$ channels, and both Ni-$d$ and O-$p$ states, respectively.}
% \label{tab:bond_lengths}
% \begin{ruledtabular}
% \begin{tabular}{ c | c | c | c | c }
% & & & PBE+$U$+$J$ & PBE+$U$+$J$ \\
% & Experiment & PBE & {Ni-$d$:} $U$,$J$ = 5,1 eV & {Ni-$d$:} $U$,$J$ = 5,1 eV \\
% & \cite{exp-LiNiPO4} & & {O-$p$:} $U$,$J$ =0 eV & {O-$p$:} $U$,$J$=9,1.5 eV \\
% \hline
% Mean Ni-O & 2.086 & 2.099 & 2.108 & 2.095 \\
% bond length (\angstrom) & $\pm$ 0.044 & $\pm$ 0.037 & $\pm$ 0.039 & $\pm$ 0.043 \\
% \end{tabular}
% \end{ruledtabular}
% \end{table*}

While the addition of Hubbard and Hunds parameters go some way to addressing the canting angle of \ce{LiNiNO4}, introducing these terms can also alter the geometry of the system. To explore this effect, we performed structural relaxations of the system with various combinations of Hubbard and Hund's corrections. In each of these structural relaxation calculations, a maximum force tolerance was used of 10 meV/$\angstrom$. All runs included spin-orbit coupling, and were constrained to the 7.8 degrees experimentally observed canting angle.

Table~\ref{tab:cellsize_UJ_LiNiPO4} lists the optimized unit cell parameters and volume, compared with the experimentally measured geometry \cite{exp-LiNiPO4}. For both the PBE+$U_\mathrm{eff}$ and PBE+$U$+$J$ schemes, adding corrections to the Ni-$d$ space worsens the geometry relative to the uncorrected PBE geometry (as earlier observed by Zhou and co-workers \cite{cococcioni-ceder-ldau-redox}. However, the further addition of corrections to the O-$p$ subspace reduces the errors by three-fold, resulting in geometries that are closest to experiment. This is similar to observations in other studies when applying corrections to O-$p$ subspaces \cite{LinscottEtAl}. We note that applying a +$J$ correction to non-magnetic O-$p$ states is unconventional. However, it should be stressed that the projected magnetic moments on \ce{LiNiPO4} remain just below 0.01 $\mu_B$, with and without on-site corrections to O-$2p$ states. Meanwhile, we can see that adding a $+J$ parameter does not significantly alter the cell parameters.

The Hubbard $U$ and Hund $J$ values used in this study of \ce{LiNiPO4} include those calculated using linear response, which are approximations of the values that are reported in Table \ref{tab:computed_uj_Ni-d_LiNiPO4}. Additionally, we tested the Ni-$d$ $U$/$J$ values used in Ref.~\onlinecite{GGAplusUJ-LiNiPO4}, in order to compare with previous computational studies of the magnetic structure of \ce{LiNiPO4}.

\subsubsection{Discussion on TM-O bond length versus \emph{U}, \emph{J}, and \emph{V} corrections}
Table \ref{tab:cellsize_UJ_LiNiPO4} also presents the change in mean Ni-O bond length between nearest-neighbor pairs for various on-site corrections. For the Ni-O bond length it is the same story as for the cell parameters: applying $U$ and $J$ to the Ni-$d$ sites worsens the results relative to the PBE result, but by applying corrections to the O-$p$ channels we obtain bond lengths that are in closer agreement with experiment. In Ref. \onlinecite{LinscottEtAl}, some of us attempted to rationalize this trend in the computed bond length between transition metal species and oxygen anions and how it improves with the introduction of corrections to the O-$p$ subspace \cite{LinscottEtAl}. We suggested that when $+U$ is added to the Ni-$d$ subspace the resulting potential shift disrupts hybridization between the Ni-$d$ and O-$p$ orbitals, weakening the bonding between these two elements (and thus leading to bond lengthening). By applying corrections to the O-$p$ re-aligns these two subspaces and allows them to ``re-hybridize".

\begin{figure*}
    \centering
    \includegraphics[width=0.98\linewidth]{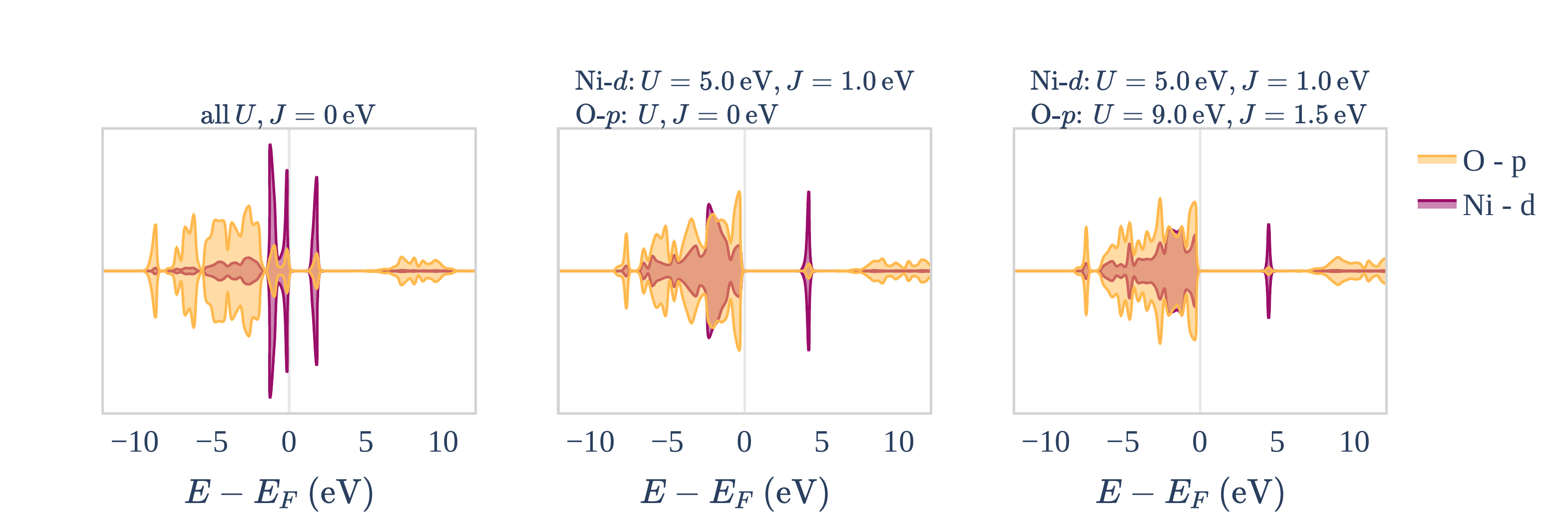}
    \caption{Projected electronic density of states for \ce{LiNiPO4} (calculated using experimental unit cell \cite{exp-LiNiPO4}) without Hubbard or Hund corrections applied, as well as $+U$ and $+J$ applied to Ni-$d$ channels, and both Ni-$d$ and O-$p$ states, respectively.}
    \label{fig:fig_dos}
\end{figure*}

In an attempt to more thoroughly explore this reasoning, Figure~\ref{fig:fig_dos} provides a comparison for the projected density of states (DOS) of \ce{LiNiPO4} for PBE and PBE+$U$+$J$ (with and without corrections to O-$p$). It is difficult to discern re-hybridization from DOS plots alone.

% Therefore, in future studies on the effect of on-site corrections to hybridization, it will behoove researchers to utilize techniques that involve the projection onto a localized, orbital-type, basis \cite{vaugier-hubbard-2012}, or employ hybridization function, which has been explored in the work of Herper et al. \cite{herper_combining_2017}.

% However, we can argue that by observing the peaks of the DOS plots for Ni-$d$ and O-$p$, there appears to be a larger correlation between the peaks of the profiles for Ni-$d$ and O-$p$ for Figure \ref{fig:fig_dos}a and Figure \ref{fig:fig_dos}c compared with Figure \ref{fig:fig_dos}b. This would confirm a restoration of the $p$-$d$ hybridization by applying on-site corrections to O-$p$. In a subsequent study, we plan to confirm the effect of $U$ and $J$ on $p$-$d$ hybridization by projecting the plane wave basis onto a localized basis set, which can be achieved in an orbital-like manner using maximally localized Wannier functions (MLWFs).

Without an explicit quantification of hybridization effects, we have added a derivation in the Supplementary Information that presents a mathematical expression of the forces acting on ions due to +$U$+$V$ corrections. This result is an extension of the theory put forth by Matteo Cococcioni in Chapter 4, Section 4.1 of Ref.~\onlinecite{corr-el-book-2012-ch4}. We argue that in quantifying the forces on TM-O bond lengths due to on-site corrections, it is possible to show that the force contributions due to both +$U^\gamma$ and +$V^{\gamma \gamma'}$ can, and should, be treated on the same footing, where $\gamma$ and $\gamma'$ correspond to atomic sites. It isn't possible to definitively state the comparative magnitude, or sign, of these force contributions without additional calculations or simplifications based on physical intuition. However, the result suggests that the forces on TM-O  bond-length due to O-$p$ $U$ values will have a comparative magnitude to the forces due to inter-site Coulomb corrections due to +$V$.

In the Supplementary Information, we further hypothesize the sign of these force contributions, starting from a DFT geometry-optimized structure without on-site corrections. Using these assumptions, which are based on computational trends in bulk TMOs, we conclude that both applying a +$U$ correction to the O-$p$ manifold and a +$V$ between TM and O states combine to mitigate the overestimation of TM-O bond length that arises when only applying +$U$ to localized states around the TM species.

\section{Conclusions}

This study provides a high-throughput \texttt{atomate} framework for calculating Hubbard $U$ and Hund's $J$ values. %$U_{\text{eff}}$ values can be calculated via the non-spin polarized response to be consistent with the Dudarev formalism \cite{Dudarev-OG}.
Using the spin-polarized linear-response methodology \cite{LinscottEtAl}, we generated a database of $U$ and $J$ values for over two thousand transition-metal-containing materials. This enabled the creation of a ``periodic table" of $U$ and $J$ values, where for each element we observe a distribution of Hubbard $U$ and Hund's $J$ values. These distributions exhibited clustering depending on the corresponding $m_{l}$ and $n_l$ values, but these quantities alone do not prove sufficient to predict the Hubbard and Hund's parameters.

In addition to +$U$+$J$, inter-site +$V$ corrections will also contribute to electronic properties. In order to investigate inter-site screening effects on the resulting $U$/$J$ values, we performed a small supercell scaling study for the full screening linear response analysis for \ce{NiO}, in addition to the conventional, atom-wise, screening. This exploration can be found in the Supplementary Information, and the details of the full screening matrix inversion can be found in Appendix \ref{sec:inversion_schemes}. We found that the full matrix inversion is much more sensitive to the size of the unit cell compared to the conventional, atom-wise screening. The theoretical reasons for this phenomenon will be an interesting pursuit for future studies, in addition to the effect of the corresponding $V^{\gamma \gamma'}$ values on the \DFTUpJp{}+$V$ ground-state. Currently, VASP does not have +$V$ corrections implemented.

% \textcolor{red}{After we realized that there were potentially issues with the full inversion results, are there any conclusions we can draw on this front?}

% Through our analysis, we found little difference between the atom-wise and full screening inversion methods, with the exceptions of oxygen $p$-electron responses. Therefore, in many cases, the atom-wise method should be sufficient for calculating $U$ and $J$ values, which precludes the need for accounting for inter-site screening effects at larger computational expense. %Compared with the ``simple" screening formula, in general the ``scaled" formula for computing Hund $J$ values had a lower associated uncertainty. 

In order to test the validity of the linear response implementation, we explored the spin-canting noncollinear magnetic structure and unit cell shape of \ce{LiNiPO4}, and compare the results with previous experimental \cite{exp-LiNiPO4} and computational \cite{GGAplusUJ-LiNiPO4, cococcioni-ceder-ldau-redox} studies. Similarly to Bousquet and Spaldin \cite{GGAplusUJ-LiNiPO4}, we observed that the computed stable canting angle was less than 50\% of the experimentally measured canting angle of nickel magnetic moments in olivine \ce{LiNiPO4}, for Ni-$d$ Hund $J$ values up to 2 eV. We also observed a large sensitivity to the canting angle and Hund $J$ values. This confirms that Hund $J$ values are crucial for exploring the properties of transition metal oxides which exhibit a noncollinear magnetic structure. In addition to the canting structure of \ce{LiNiPO4}, we also presented the relaxed unit cell shape for various Hubbard $U$ and Hund $J$ corrections. While applying a +$U$+$J$ correction to Ni-$d$ resulted in increased disagreement with experimentally measured unit cell parameters \cite{cococcioni-ceder-ldau-redox}, applying an on-site Hubbard/Hund correction to O-$p$ occupancies greatly improved the agreement of unit cell shape with experiment \cite{exp-LiNiPO4}. This finding reinforces the importance of including a +$U$+$J$ correction to oxygen sites in order to resolve the accurate bonding behavior between transition metal species and neighboring oxygen atoms.

%The spin-polarized LR method exhibits an additional scalability over the non-spin-polarized counterpart. We observed that for the \ce{LiNiPO4} case study, the Ni-$d$ Hubbard $U$ values changed less than 20 meV for a larger supercell, compared with the single parent unit cell. This result contrasts with the findings of Cococcioni and others, who computed the convergence of Hubbard $U$ values only at much larger supercells due to self-screening due to periodic boundary conditions \cite{PhysRevB.71.035105}. Further convergence studies will be needed to confirm this improved finite-size scaling as a result of the screening between spin channels.

% Future studies may find it important to explore how these custom-computed $U_{\text{eff}}$ and/or $U$ and $J$ values influence additional GGA+$U$+$J$ derived ground-state material properties on high-throughput computational scales. In particular, a possible focus could be on the magnetic properties of transition metal oxides. Namely, an important subsequent study would be one that centers around elucidating the extent to which the computed collinear \cite{horton-high-throughput-2019} and noncollinear ground-state of magnetic materials changes with applied $U$ and $J$ values. This will also include the magnitude of magnetic moments, which has been shown to be highly sensitive to $U$ and $J$ values \cite{LinscottEtAl}.

\section*{Acknowledgements}

The authors would like to thank Professor Matteo Cococcioni for his helpful correspondence over email, and for addressing questions on the original linear response methodology. G.M. acknowledges support from the Department of Energy Computational Science Graduate Fellowship (DOE CSGF) under grant DE-SC0020347. E.L. acknowledges support from the Swiss National Science Foundation (SNSF) under grant 200021-179138. Computations in this paper were performed using resources of the National Energy Research Scientific Computing Center (NERSC), a U.S. Department of Energy Office of Science User Facility operated under contract no. DE-AC02-05CH11231. Expertise in high-throughput calculations, data and software infrastructure was supported by the U.S. Department of Energy, Office of Science, Office of Basic Energy Sciences, Materials Sciences and Engineering Division under Contract DE-AC02-05CH11231: Materials Project program KC23MP.

\section*{CRediT Taxonomy}

We highlight the author contributions to this study using the CRediT taxonomy. \textbf{Guy C. Moore}: Conceptualization, Methodology, Software, Validation, Formal analysis, Investigation, Data Curation, Writing - Original Draft, Writing - Review \& Editing, Visualization \textbf{Matthew K. Horton}: Conceptualization, Software, Validation, Investigation, Writing - Review \& Editing, Visualization, Project administration \textbf{Alexander M. Ganose}: Software, Writing - Review \& Editing \textbf{Edward Linscott}: Methodology, Validation, Formal analysis, Investigation, Writing - Review \& Editing \textbf{David D. O'Regan}: Methodology, Validation, Formal analysis, Investigation, Writing - Review \& Editing \textbf{Martin Siron}: Software, Writing - Review \& Editing  \textbf{Kristin A. Persson}: Writing - Review \& Editing, Supervision, Project administration.

% \newpage

\appendix
\section{Screening matrix inversions}
\label{sec:inversion_schemes}
\noindent Below are the matrix representations of the response matrices at each level of screening outlined by Linscott and others for a system with two Hubbard sites \cite{LinscottEtAl}.

\hspace{2ex}

\noindent Point-wise $1\times1$ screening:
\begin{align}
    \chi^{-1} &=
    \begin{pmatrix} 
        1/\chi_{11} & 0 \\ 0 & 1/\chi_{22} \\
    \end{pmatrix}
\end{align}
Atom-wise (conventional) $2\times2$ screening:
\begin{align}
    \chi^{-1} &=
    \begin{pmatrix} 
        \chi_{11} & \chi_{12} \\ \chi_{21} & \chi_{22} \\
    \end{pmatrix}^{-1}
\end{align}
We can extend this formalism to the multiple site (multi-site) responses by considering the response matrix for two sites, $\chi_{ij}$, where $i$ and $j$ are the site indices. 

\noindent Point-wise screening:
\begin{align}
    \chi^{-1} &=
    \begin{pmatrix} 
        \begin{pmatrix} 
            1/\chi^{\uparrow \uparrow}_{11} & 0 \\
            0 & 1/\chi^{\downarrow \downarrow}_{11} \\
        \end{pmatrix} & \text{\Large 0} \\ \text{\Large 0} & 
        \begin{pmatrix} 
            1/\chi^{\uparrow \uparrow}_{22} & 0 \\
            0 & 1/\chi^{\downarrow \downarrow}_{22} \\
        \end{pmatrix} \\
    \end{pmatrix} 
\end{align}
Atom-wise (conventional) screening:
\begin{align}
    \chi^{-1} &=
    \begin{pmatrix} 
        \begin{pmatrix} 
            \chi^{\uparrow \uparrow}_{11} & \chi^{\uparrow \downarrow}_{11} \\
            \chi^{\downarrow \uparrow}_{11} & \chi^{\downarrow \downarrow}_{11} \\
        \end{pmatrix}^{-1} & \text{\Large 0} \\ \text{\Large 0} & 
        \begin{pmatrix} 
            \chi^{\uparrow \uparrow}_{22} & \chi^{\uparrow \downarrow}_{22} \\
            \chi^{\downarrow \uparrow}_{22} & \chi^{\downarrow \downarrow}_{22} \\
        \end{pmatrix}^{-1} \\
    \end{pmatrix}
\end{align}
Full screening:
\begin{align}
    \chi^{-1} &=
    \begin{pmatrix} 
        \chi^{\uparrow \uparrow}_{11} & \chi^{\uparrow \downarrow}_{11} &
        \chi^{\uparrow \uparrow}_{12} & \chi^{\uparrow \downarrow}_{12} \\
        \chi^{\downarrow \uparrow}_{11} & \chi^{\downarrow \downarrow}_{11} & 
        \chi^{\downarrow \uparrow}_{12} & \chi^{\downarrow \downarrow}_{12} \\
        \chi^{\uparrow \uparrow}_{21} & \chi^{\uparrow \downarrow}_{21} &
        \chi^{\uparrow \uparrow}_{22} & \chi^{\uparrow \downarrow}_{22} \\
        \chi^{\downarrow \uparrow}_{21} & \chi^{\downarrow \downarrow}_{21} & 
        \chi^{\downarrow \uparrow}_{22} & \chi^{\downarrow \downarrow}_{22} \\
    \end{pmatrix}^{-1}
\end{align}
We note that it is important when performing a linear response calculation to construct a $2N \times 2N$ response matrix where $N$ is the number of Hubbard sites (or $N \times N$ in the case of non-spin-polarized linear response). For bulk systems often several Hubbard sites will be equivalent, and one can save computational time by performing linear response calculations for the set of inequivalent sites, and then populating the response matrix for all equivalent Hubbard-site pairs.
% \clearpage
% \newpage

\section{Post-processing \& uncertainty quantification}

In order to extract the response matrices from the raw DFT data, curve fitting was performed using a least-squares polynomial fit implemented in \texttt{numpy} \cite{numpy}. The uncertainty associated with each computed slope was obtained from the covariance matrix produced as a result of the least-squares fit. These uncertainty values were then utilized to determine the errors associated with the Hubbard $U$ and Hund $J$ values. The error quantification was performed by computing the propagation of uncertainty based on the Jacobian of each scaling formula for Hubbard $U$ and Hund $J$. This method for error propagation is general to multiple levels of screening between spin, site, and orbital responses.

We begin by considering the following screening matrix introduced in Equation \ref{eq:fmatrix}, from which Hubbard $U$ and Hund $J$ values are derived \cite{LinscottEtAl}
\begin{align*}
    f_{ij} = \left(\chi_0^{-1} - \chi^{-1}\right)_{ij}
\end{align*}
Derivatives of the $\chi^{-1}$ matrix with respect to individual $\chi_{kl}$ can be obtained by the following relation:
\begin{align}
    \frac{\partial}{\partial \chi_{kl}} \left(\chi^{-1}\right) &= - \chi^{-1} \left( \frac{\partial}{\partial \chi_{kl}} \chi \right) \chi^{-1} \nonumber \\
    & \text{where } \frac{\partial}{\partial \chi_{kl}} \left\{ \chi \right\}_{ij} = 
    \begin{cases}
    1 & \text{if } kl=ij \\
    0 & \text{otherwise}
    \end{cases} \nonumber \\
    \frac{\partial}{\partial \chi_{kl}} \left\{ \chi^{-1} \right\}_{ij} &= - \left\{\chi^{-1}\right\}_{ik} \left\{ \chi^{-1} \right\} _{lj}
    \label{eq:inv_deriv}
\end{align}
Using this fact, it is possible to obtain the full Jacobian of $f$ with respect to response $\chi$ matrices which can be used to obtain the covariance uncertainty matrix associated with the elements of $f_{ij}$, to a first-order expansion of $f_{ij}$ \cite{ochoa-covariance-nodate}
\begin{align}
    \bm \Sigma_{f} &= \bm J_{\chi_0} \bm \Sigma_{\chi_0} \bm J_{\chi_0} ^{T} + \bm J_{\chi} \bm \Sigma_{\chi} \bm J_{\chi} ^{T}
\end{align}
where $\bm \Sigma_{f}$ is a $N^2 \times N^2$ matrix ($f$ is $N\times N$). Each element of $\bm \Sigma_{f}$, $\left\{ \Sigma_{f}\right\}_{ij,kl}$, corresponds to the covariance between $f_{ij}$ and $f_{kl}$ matrix elements. $\bm \Sigma_{\chi}$ and $\bm \Sigma_{\chi_0}$ are the covariance matrices for each $\left\{ \chi \right\}_{kl}$ and $\left\{\chi_0\right\}_{kl}$, and the diagonal elements are populated using the squared uncertainty values associated with the slopes fit to the response data. In addition, $\bm J_{\chi}$ and $\bm J_{\chi_0}$ are the symbolically derived Jacobians corresponding to each response value, as proposed in Equation \ref{eq:inv_deriv}. Assuming that the individual elements of $\chi$ and $\chi_0$ are independent, we can assume that $\bm \Sigma$ covariance matrices are diagonal in order to make the following simplification:
\begin{align}
    \sigma^2 (f_{ij}) &= \sum_{kl} \left(\frac{\partial}{\partial \left\{\chi_0\right\}_{kl}} f_{ij}\right)^2 \sigma^2 (\left\{\chi_0\right\}_{kl}) \nonumber \\
    & \quad + \sum_{kl} \left(\frac{\partial }{\partial \left\{\chi\right\}_{kl}} f_{ij} \right)^2 \sigma^2 (\left\{\chi\right\}_{kl}), \label{eq:s2_f}
\end{align}
where $\sigma^2 (f_{ij})$, $\sigma^2 (\left\{\chi_0\right\}_{ij})$, and $\sigma^2 (\left\{\chi\right\}_{ij})$ correspond to the diagonal elements of $\bm \Sigma_{f}$, $\bm \Sigma_{\chi_0}$,  and $\bm \Sigma_{\chi}$, respectively.

With the established expression for the uncertainty values of $f$ in Equation \ref{eq:s2_f}, we can express the squared uncertainty of $U$, for an atomic site $\gamma$, in the next level of uncertainty propagation,
\begin{align}
    \sigma^2 (U^\gamma) &= \sum_{\sigma,\sigma'} \left( \frac{\partial}{\partial f^{\sigma \sigma'}_{\gamma \gamma}} G_U(f_{\gamma \gamma})\right)^2 \sigma^2 (f^{\sigma \sigma'}_{\gamma \gamma}). \label{eq:s2_u}
\end{align}
Equation \ref{eq:s2_u} can be extended to an expression of the squared uncertainty of Hund $J$, where $G_U$ and $G_J$ are functions of 2$\times$2 sub-matrices along the diagonal of $f$, as introduced in Equation \ref{eq:fmatrix}, and depend on the different scaling schemes introduced in Ref.~\onlinecite{LinscottEtAl}.

\newpage
% \subsection{Sample statistics}

\section{Details of the data behind the periodic tables}
\label{sec:sample_details}

\begin{table}[H]
\caption{The mean and standard deviation ($\sigma$) in the $U$ and $J$ parameters used in the periodic tables of Figure~\ref{fig:ptable}, alongside the number of samples $N$.}
% \label{tab:}
\begin{ruledtabular}
\begin{tabular}{lrrrrrr}
\multirow{2}{*}{element} & \multicolumn{3}{c}{$U$} & \multicolumn{3}{c}{$J$} \\
& \multicolumn{1}{c}{mean}
& \multicolumn{1}{c}{$\sigma$}
& \multicolumn{1}{c}{$N$}
& \multicolumn{1}{c}{mean}
& \multicolumn{1}{c}{$\sigma$}
& \multicolumn{1}{c}{$N$} \\
\hline
 Mn    & 4.710 & 0.707 &  94 &  0.575 & 0.157 &  97 \\
 Fe    & 4.545 & 0.674 &  78 &  0.437 & 1.137 & 122 \\
  V    & 3.909 & 0.404 &  68 &  0.849 & 0.538 & 108 \\
 Cu    & 7.590 & 0.728 &  51 &  1.117 & 1.083 &  71 \\
 Cr    & 2.982 & 0.464 &  51 &  0.557 & 0.089 &  61 \\
 Nb    & 0.529 & 0.107 &  47 &  0.193 & 0.054 &  39 \\
 Ti    & 4.737 & 0.428 &  45 &  0.705 & 0.861 &  62 \\
 Ta    & 3.688 & 0.130 &  34 &  0.628 & 0.079 &  37 \\
  W    & 1.846 & 0.213 &  33 &  0.385 & 0.045 &  33 \\
 Co    & 5.237 & 0.566 &  33 &  0.807 & 1.361 &  46 \\
 Ag    & 2.830 & 0.606 &  26 &  0.703 & 0.131 &  24 \\
 Re    & 0.598 & 0.172 &  26 &  0.255 & 0.089 &  27 \\
 Ni    & 5.847 & 0.704 &  25 &  0.589 & 0.320 &  33 \\
 Zr    & 4.382 & 0.269 &  24 &  0.740 & 0.069 &  23 \\
 Mo    & 2.431 & 0.230 &  21 &  0.520 & 0.220 &  28 \\
 Hg    & 0.620 & 0.226 &  21 &  0.288 & 0.271 &  22 \\
 Cd    & 0.350 & 0.327 &  19 &  0.609 & 0.614 &   5 \\
 Sc    & 2.506 & 0.210 &  16 &  0.543 & 0.070 &  16 \\
  Y    & 4.704 & 0.393 &  15 &  0.825 & 0.179 &   5 \\
 Pt    & 1.673 & 0.318 &  13 &  0.322 & 0.201 &  14 \\
 Os    & 1.855 & 0.448 &  10 &  0.361 & 0.087 &  11 \\
 Ru    & 2.972 & 0.548 &  10 &  0.504 & 0.292 &  24 \\
 Lu    & 0.449 & 0.065 &   9 &  0.292 & 0.072 &   8 \\
 Pd    & 3.608 & 0.407 &   8 &  0.620 & 0.106 &  10 \\
 Hf    & 3.733 & 0.299 &   8 &  0.812 & 0.122 &   8 \\
 Au    & 1.186 & 0.256 &   7 &  0.484 & 0.165 &   8 \\
 Zn    & 0.530 & 0.795 &   5 & -0.105 & 0.433 &  17 \\
 Rh    & 1.616 & 0.201 &   5 &  0.406 & 0.065 &   5 \\
 Ir    & 1.868 & 0.288 &   5 &  0.352 & 0.283 &  14 \\
 Tc    & 2.956 & 0.100 &   3 &  0.980 & 1.247 &   9 \\
 Total &       &       & 810 &        &       & 987
\end{tabular}
\end{ruledtabular}
\end{table}

\nocite{*}
\bibliography{main}

% \nolinenumbers

\end{document}

% \preprint{APS/123-QED}

\title{Supplementary information: \\ \textit{High-throughput determination of Hubbard $U$ and Hund $J$ values for transition metal oxides via linear response formalism}}% Force line breaks with \\

\author{
Guy C. Moore,
Matthew K. Horton,
Alexander M. Ganose, \\
Edward Linscott,
David D. O'Regan, \\
Martin Siron,
Kristin A. Persson}

\date{\today}% It is always \today, today,

\maketitle

\newpage
\section{\ce{LiNiPO4} response plots}

The $2 \times 2$ spin-polarized response for Ni-$d$ states is shown in Figure \ref{fig:response_plots_A_LiNiPO4}. In this figure, each subplot corresponds to a different component to the spin-channel response tensor, $\chi^{\sigma \sigma'}$, where $\sigma, \sigma' = \uparrow \text{or} \downarrow$.

\begin{figure}[H]
    \centering
    \includegraphics[height=0.75\linewidth]{si_images/UJpaper_response_full.pdf}
    \caption{Linear response results for Ni-$d$ within \ce{LiNiPO4}; Response data and linear fit shown in magenta for self-consistent (SCF) and orange for non-self-consistent (NSCF) results; $\chi^{\uparrow \uparrow}$, $\chi^{\uparrow \downarrow}$,  $\chi^{\downarrow \uparrow}$, and $\chi^{\downarrow \downarrow}$ responses corresponding to elements of $2 \times 2$ response matrix}
    \label{fig:response_plots_A_LiNiPO4}
\end{figure}

\section{NiO test case}

\subsection{Supercell scaling test \& Pseudopotential comparison}
\label{sec:scaling_pp}

In this preliminary test, we perform a supercell scaling study for NiO using the antiferromagnetic (AFM) primitive cell, with lattice vectors $\bm a = [-2.02, +0.00, +2.02]$ \angstrom, $\bm b = [-2.02, +2.02, +0.00]$ \angstrom, and $\bm c = [-4.04, -2.02, -2.02]$ \angstrom. In all calculations, we test the PBE GGA functional, and compare the various PAW PBE 52 pseudopotentials for nickel and oxygen, namely: \texttt{Ni\_pv}, \texttt{Ni}, \& \texttt{Ni\_GW}, as well as \texttt{O} \& \texttt{O\_GW}, respectively. While we found that supercell size did not change the $U$ values outside of uncertainty, the type of pseudopotential (PP) used had a very significant effect on the U/J values that resulted. While this isn't all together surprising, this demonstration should serve to underscore the lack of transferability of on-site corrections, which are calculated using the same DFT functional, but a different PP.

\begin{table}[H]
\centering
\caption{Ni-$d$ Hubbard $U$ scaled atom-wise (in eV)}
% \label{tab:}
\begin{tabular}{ c | c c c}
Supercell & \multicolumn{3}{c}{PAW PBE 52 pseudopotential} \\
& \texttt{Ni\_pv} & \texttt{Ni} & \texttt{Ni\_GW} \\
\hline
$1 \times 1 \times 1$ & 6.10 $\pm$ 0.19 & 5.27 $\pm$ 0.12 & 1.50 $\pm$ 0.03 \\
$2 \times 1 \times 1$ & 5.92 $\pm$ 0.17 & 5.21 $\pm$ 0.13 & 1.58 $\pm$ 0.03 \\
$2 \times 2 \times 1$ & 6.04 $\pm$ 0.17 & 5.24 $\pm$ 0.15 & 1.57 $\pm$ 0.03 \\
$2 \times 2 \times 2$ & 6.15 $\pm$ 0.18 & 5.19 $\pm$ 0.14 & 1.56 $\pm$ 0.03 \\
$4 \times 2 \times 2$ & 6.09 $\pm$ 0.16 & - & - \\
$4 \times 4 \times 4$ & 6.02 $\pm$ 0.16 & - & - \\
\end{tabular}
\end{table}

\begin{table}[H]
\centering
\caption{Ni-$d$ Hund $J$ scaled atom-wise (in eV)}
% \label{tab:}
\begin{tabular}{ c | c c c}
Supercell & \multicolumn{3}{c}{PAW PBE 52 pseudopotential} \\
& \texttt{Ni\_pv} & \texttt{Ni} & \texttt{Ni\_GW} \\
\hline
$1 \times 1 \times 1$ & 0.846 $\pm$ 0.032 & 0.757 $\pm$ 0.023 & 0.315 $\pm$ 0.009\\
$2 \times 1 \times 1$ & 0.709 $\pm$ 0.030 & 0.668 $\pm$ 0.025 & 0.269 $\pm$ 0.008 \\
$2 \times 2 \times 1$ & 0.733 $\pm$ 0.029 & 0.670 $\pm$ 0.025 & 0.267 $\pm$ 0.008 \\
$2 \times 2 \times 2$ & 0.700 $\pm$ 0.029 & 0.662 $\pm$ 0.025 & 0.261 $\pm$ 0.009 \\
$4 \times 2 \times 2$ & 0.710 $\pm$ 0.027 & - & - \\
$4 \times 4 \times 4$ & 0.707 $\pm$ 0.028 & - & - \\
\end{tabular}
\end{table}

\begin{table}[H]
\centering
\caption{O-$p$ scaled atom-wise}
% \label{tab:}
\begin{tabular}{ c | c c | c c}
Supercell & \multicolumn{4}{c}{PAW PBE 52 pseudopotential} \\
& \multicolumn{2}{c|}{\texttt{O}} & \multicolumn{2}{c}{\texttt{O\_GW}} \\
%\cline{2-5}
& $U$ (eV) & $J$ (eV) & $U$ (eV) & $J$ (eV) \\
\hline
$1 \times 1 \times 1$ & 9.1868 $\pm$ 0.1812 & 1.2290 $\pm$ 0.0345 & 8.61 $\pm$ 0.04 & 1.133 $\pm$ 0.013 \\
$2 \times 1 \times 1$ & 9.7752 $\pm$ 0.1838 & 1.2290 $\pm$ 0.0345 & 8.78 $\pm$ 0.04 & 1.215 $\pm$ 0.033 \\
$2 \times 2 \times 1$ & 9.7364 $\pm$ 0.2848 & 1.3587 $\pm$ 0.0410 & 8.72 $\pm$ 0.06 & 1.321 $\pm$ 0.053 \\
$2 \times 2 \times 2$ & 9.6771 $\pm$ 0.1854 & 1.2681 $\pm$ 0.0410 & 8.52 $\pm$ 0.07 & 1.424 $\pm$ 0.065 \\
\end{tabular}
\end{table}

\subsection{Full matrix inversion}

In this short scaling study on NiO, we perform the full inversion on the primitive cell, and a supercell with twice as many atoms, as shown in Table \ref{tab:full_scaling}. These values are calculated with the screening between all atoms included in the matrix inversion, i.e. $8 \times 8$ matrix in the $1 \times 1 \times 1$ cell ($2\times$ spin-channels $\times$ ($2 \times$ Ni atoms + $2 \times$ O atoms) = 8 sites), and $16 \times 16$ matrix in the $2 \times 1 \times 1$ cell.

\begin{table}[H]
\centering
\caption{Ni-$d$ scaled full (\texttt{Ni\_pv} PAW PBE 52 pseudopotential)}
\label{tab:full_scaling}
\begin{tabular}{ c | c c }
Supercell & $U$ (eV) & $J$ (eV) \\
\hline
$1 \times 1 \times 1$ & 4.445 $\pm$ 0.134 & 1.0173 $\pm$ 0.0398 \\
$2 \times 1 \times 1$ & 5.311 $\pm$ 0.095 & 0.9178 $\pm$ 0.0240 \\
\end{tabular}
\end{table}

\noindent It is evident that these values change significantly between the two cell sizes - by nearly 1 eV. Therefore, we can conclude that for the full inversion, it is clear that more rigorous supercell scaling must be performed. This is a question that we hope will be addressed in future studies, which are interested in using this framework for obtaining $V^{\gamma \gamma'}$ inter-site exchange values relevant for DFT+$U$+$V$.

\subsection{Self-consistent U \& J experiment}

In this section, we performed a manual ``self-consistency" cycle for the NiO test case. The first iteration uses the relaxed primitive cell from using the MP default $U_{eff}$  for Ni-$d$ of 6.2 eV. For each geometric optimization step, we use an SCF energy convergence of $1.0^{-4}$ eV (\texttt{EDIFF}), which corresponds to a ionic relaxation tolerance of $1.0^{-3}$ eV (\texttt{EDIFFG}). As is evident from the table below, these tolerances were sufficient in order to achieve self-consistent $U$ and $J$ values for Ni-$d$ and O-$p$, within their corresponding uncertainties. The smaller uncertainty values are due to the fact that we use seventeen evaluations for the linear response analysis, instead of nine in the previous section. For this system, it is apparent that self-consistent on-site corrections are achieved within the first iteration. However, this may not be true for more complex oxides, in which the rotation of oxygen polyhedra play a more crucial role.

\begin{table}[H]
\centering
\caption{Cell relaxation \& linear response iterations}
% \label{tab:}
\begin{tabular}{ c | c c c c c}
Iter- & \multicolumn{2}{c}{Ni-$d$:} & \multicolumn{2}{c}{O-$p$:} & \\
ation & U (eV) & J (eV) & U (eV) & J (eV) & Lattice vectors (\angstrom) \\
\hline
& & & & & $\bm a = [-2.34, +0.00, +2.37]$ \\
1 & 5.831 $\pm$ 0.064 & 0.422 $\pm$ 0.021 & 10.326 $\pm$ 0.068 & 1.103 $\pm$ 0.008 & $\bm b = [-2.34, +2.36, +0.00]$ \\
& & & & & $\bm c = [-4.67, -2.35, -2.36]$ \\
& & & & & \\
& & & & & $\bm a = [-2.23, +0.00, +2.23]$ \\
2 & 5.884 $\pm$ 0.070 & 0.551 $\pm$ 0.021 & 9.755 $\pm$ 0.053 & 1.125 $\pm$ 0.007 & $\bm b = [-2.23, +2.23, +0.00]$ \\
& & & & & $\bm c = [-4.45, -2.23, -2.23]$\\
& & & & & \\
& & & & & $\bm a = [-2.23, +0.00, +2.23]$ \\
3 & 5.881 $\pm$ 0.070 & 0.547 $\pm$ 0.021 & 9.788 $\pm$ 0.054 & 1.124 $\pm$ 0.007 & $\bm b = [-2.23, +2.23, +0.00]$ \\
& & & & & $\bm c = [-4.45, -2.23, -2.23]$ \\
& & & & & \\
\end{tabular}
\end{table}

\subsection{Electron localization function}

In our study of NiO, we apply the electron localization function (ELF) to the system, as defined by Silvi et al. \cite{silvi-classification-1994} and implemented in VASP, which is written to the ELFCAR output file. Other studies have shown that the ELF provides insight into locations of high self-interaction error (SIE) \cite{hao_using_2014}. In Figures \ref{fig:elf_a} - \ref{fig:elf_c}, we provide a comparison over the visualized isosurfaces of the ELF for various levels of on-site corrections. We continue to focus on the antiferromagnetic primitive cell of NiO studied in Section \ref{sec:scaling_pp}.

\begin{figure}[H]
    \centering
    \subfloat[\centering All $U,J = 0.0$ eV]{
    \includegraphics[width=0.32\linewidth]{si_images/ELFCAR_noUJ.png}
    \label{fig:elf_a}
    }
    \subfloat[\centering Ni-$d$: $U,J=6.3,0.8$ eV \protect\newline O-$p$: $U,J=0.0$ eV]{
    \includegraphics[width=0.32\linewidth]{si_images/ELFCAR_UJ_Ni.png}
    \label{fig:elf_b}
    }
    \subfloat[\centering Ni-$d$: $U,J=6.3,0.8$ eV \protect\newline O-$p$: $U,J=10.0,1.5$ eV]{
    \includegraphics[width=0.32\linewidth]{si_images/ELFCAR_UandJ_Ni_and_O.png}
    \label{fig:elf_c}
    }
    \caption{Visualization of electron localization function iso-surfaces in NiO cell (a) without Hubbard or Hund corrections applied, as well as (b) +U and +J applied to Ni-d channels, and (c) both Ni-d and O-p states, respectively. The two nickel atoms are located in the center and corners of the unit cell. The isosurface levels are at 0.142 and 0.146.}
\end{figure}

\noindent Applying the +$U$+$J$ to all sites in this case appears to have the strongest effect on the localization around the O-$p$ states. It is interesting to note that it appears that in Figure \ref{fig:elf_b}, applying on-site corrections to Ni-$d$ sites appears to have a stronger effect on the localization function surrounding the oxygen atoms than the nickel atoms. While these visualizations are interesting to observe, it is difficult to conclude why a larger $U$ value is needed for O-$p$ sites.

\section{Occupations calculated from DFT (no on-site corrections) versus guessed oxidation state}

In Figures \ref{fig:Mn_oxi_m}-\ref{fig:Ni_oxi_m} we provide scatter plots of the locally projected $d$ ($l=2$) component of the magnetic moments, $|m$-$d|$, versus oxidation state guessed using the bond valence method (BVM) \cite{bvm-method-paper} for a set of Mn, Fe, and Ni containing structures. Analogous plots versus total $d$ occupations, $n$-$d$, are provided in Figures \ref{fig:Mn_oxi_n}-\ref{fig:Ni_oxi_n}. We emphasize again that these local projections are obtained from VASP DFT calculations, without any on-site corrections applied. We use the BVM implemented in \texttt{pymatgen} \cite{pymatgen-github}.

\begin{figure}[H]
    \centering
    \subfloat[\centering $m$-$d$ of Mn]{{
    \label{fig:Mn_oxi_m}
    \includegraphics[width=.30\linewidth]{si_images/Mn_md_vs_oxi.png} }}
    \subfloat[\centering $m$-$d$ of Fe]{{
    \label{fig:Fe_oxi_m}
    \includegraphics[width=.30\linewidth]{si_images/Fe_md_vs_oxi.png} }}
    \subfloat[\centering $m$-$d$ of Ni]{{
    \label{fig:Ni_oxi_m}
    \includegraphics[width=.30\linewidth]{si_images/Ni_md_vs_oxi.png} }}
    %\subfloat[\centering O-$p$]{{
    %%\label{fig:}
    %\includegraphics[width=.32\linewidth]{images/} }}
    \\
    \subfloat[\centering $n$-$d$ of Mn]{{
    \label{fig:Mn_oxi_n}
    \includegraphics[width=.30\linewidth]{si_images/Mn_nd_vs_oxi.png} }}
    \subfloat[\centering $n$-$d$ of Fe]{{
    \label{fig:Fe_oxi_n}
    \includegraphics[width=.30\linewidth]{si_images/Fe_nd_vs_oxi.png} }}
    \subfloat[\centering $n$-$d$ of Ni]{{
    \label{fig:Ni_oxi_n}
    \includegraphics[width=.30\linewidth]{si_images/Ni_nd_vs_oxi.png} }}
    %\subfloat[\centering O-$p$]{{
    %%\label{fig:}
    %\includegraphics[width=.32\linewidth]{images/} }}
    \caption{Each sub-figure provides a scatter plot of the computed $m$-$d$ and $n$-$d$ site occupations for Mn, Fe, and Ni, obtained from DFT calculations (without +$U/J$). These values are plotted against the ``guessed" oxidation states using the bond valence method (BVM), as implemented in \cite{pymatgen-github}.} 
    %\label{fig:}
\end{figure}

\noindent As expected from chemical intuition, there are clear trends of $|m$-$d|$ and $n$-$d$ versus guessed oxidation. However, in a few cases, namely the trends of $|m$-$d|$ versus Mn oxidation state in Figure \ref{fig:Mn_oxi_m}, the trends of $|m$-$d|$ is clearer than that of $n$-$d$. For this reason, we provide plots of the calculated $U$ and $J$ values versus $|m$-$d|$ in the manuscript. One of the goals of this study is to lay the groundwork for a future study to explore chemical and/or physical justification for these trends.

\section{An exploration of inter-site Hubbard forces due to both +$U$ and +$V$ contributions}

\subsection{Including on-site forces due to +$V$, in addition to +$U$}

The following is based on Section 4.1 of Chapter 4, by Matteo Cococcioni, of \cite{correlated-electrons-book}, ``The LDA+U Approach: A Simple Hubbard Correction for Correlated Ground States." Our goal is to expand on Section 4.1 to include the +$V$ contributions to the Hubbard forces, in addition to +$U$, with the goal of providing a footing on which to compare the force contributions from +$U$ versus +$V$. The force acting on site $\alpha$ in spatial direction $i$ due to the +$U$+$V$ correction can be expressed as
\begin{align}
    F^{UV}_{\alpha,i} &= - \frac{\partial E_{UV}}{\partial \tau_{\alpha i}}
    = - \left( \frac{\partial E_{U}}{\partial \tau_{\alpha i}} + \frac{\partial E_{V}}{\partial \tau_{\alpha i}} \right)
\end{align}
$\tau_{\alpha i}$ is the $i$\textsuperscript{th} component of the position vector corresponding to a site indexed by $\alpha$. The partial derivatives can be expressed using the chain rule, following from Equation (20) in \cite{correlated-electrons-book}:
\begin{align}
    \frac{\partial E_{U}}{\partial \tau_{\alpha i}} &= \sum_{\gamma,m,m',\sigma} \frac{\partial E_U}{\partial n^{\gamma\gamma\sigma}_{mm'}} \frac{\partial n^{\gamma\gamma\sigma}_{mm'}}{\partial \tau_{\alpha i}} \\
    \frac{\partial E_{V}}{\partial \tau_{\alpha i}} &= \sum_{\substack{\gamma,{\gamma'},m,m',\sigma \\ \gamma \ne {\gamma'}}} \frac{\partial E_V}{\partial n^{\gamma{\gamma'}\sigma}_{mm'}} \frac{\partial n^{\gamma{\gamma'}\sigma}_{mm'}}{\partial \tau_{\alpha i}}
\end{align}
where the sum is over site indices $\gamma$ \& ${\gamma'}$, $m$ \& $m'$ quantum numbers, and $\sigma$ spin channels. From the definition of $E_{UV}$, Equation (54) in \cite{correlated-electrons-book}, we can write derivatives of the +$U$+$V$ energy with respect to occupancy numbers
\begin{align}
    \frac{\partial E_U}{\partial n^{\gamma\gamma\sigma}_{mm'}} 
    &= u^{\gamma\gamma\sigma}_{mm'} 
    = - \frac{U^\gamma}{2} \left( 2 n^{\gamma\gamma\sigma}_{m'm} - \delta_{m'm} \right) \label{eq:eu_deriv} \\
    \frac{\partial E_V}{\partial n^{\gamma{\gamma'}\sigma}_{mm'}} 
    &= v^{\gamma{\gamma'}\sigma}_{mm'}
    = - V^{{\gamma'}\gamma} n^{{\gamma'}\gamma\sigma}_{m'm} \label{eq:ev_deriv}
\end{align}
The occupancy numbers are defined as
\begin{align}
    n^{\gamma{\gamma'}\sigma}_{mm'} &= \sum_{k,v} f^\sigma_{kv} \braket{\psi^\sigma_{kv}|\phi^{\gamma'}_{m'}} \braket{\phi^\gamma_{m}|\psi^\sigma_{kv}}. \label{eq:nocc_defn}
\end{align}
in Equation (51) in \cite{correlated-electrons-book}. The derivatives of occupancy numbers w.r.t. positions $\tau_{\alpha i}$, can be expressed as 
\begin{align}
    \frac{\partial n^{\gamma{\gamma'}\sigma}_{mm'}}{\partial \tau_{\alpha i}} 
    &= \sum_{k,v} f^\sigma_{kv} \left[ 
    \frac{\partial}{\partial \tau_{\alpha i}} \left(\braket{\psi^\sigma_{kv}|\phi^{\gamma'}_{m'}}\right) \braket{\phi^\gamma_{m}|\psi^\sigma_{kv}} + 
    \braket{\psi^\sigma_{kv}|\phi^{\gamma'}_{m'}} \frac{\partial}{\partial \tau_{\alpha i}} \left( \braket{\phi^\gamma_{m}|\psi^\sigma_{kv}} \right)
    \right] \nonumber \\
    &= \sum_{k,v} f^\sigma_{kv} \left[ 
    \delta_{\alpha {\gamma'}} \braket{\psi^\sigma_{kv}|\phi^\gamma_{m}}^* \frac{\partial}{\partial \tau_{{\gamma'} i}} \left( \braket{\phi^{\gamma'}_{m'}|\psi^\sigma_{kv}}^* \right) + 
    \delta_{\alpha \gamma} \braket{\psi^\sigma_{kv}|\phi^{\gamma'}_{m'}} \frac{\partial}{\partial \tau_{\gamma i}} \left( \braket{\phi^\gamma_{m}|\psi^\sigma_{kv}} \right)
    \right] \nonumber \\
    &= \delta_{\alpha {\gamma'}} \left(\tilde n^{{\gamma'}\gamma\sigma}_{m'm}\right)^*_i + \delta_{\alpha \gamma} \left(\tilde n^{\gamma{\gamma'}\sigma}_{mm'}\right)_i \label{eq:occ_deriv}
\end{align}
Note the short-hands that we have defined so far for energy and occupancy partial derivatives in Eqn.'s \ref{eq:eu_deriv}, \ref{eq:ev_deriv}, and \ref{eq:occ_deriv}: $u^{\gamma\gamma\sigma}_{mm'}$, $v^{\gamma{\gamma'}\sigma}_{mm'}$, and $\left(\tilde n^{\gamma{\gamma'}\sigma}_{mm'}\right)_i$, respectively. The Kronecker deltas in front of each term is discussed following Equation (26) in \cite{correlated-electrons-book}, which provides some crucial insight highlighted by Cococcioni,
\begin{blockquote}
    \textit{``[T]he derivative of the atomic wave function is different from zero only in the case it is centered on the atom which is being displaced. Thus, the derivative in Eq.(23) only contributes to forces on atoms subject to the Hubbard correction. Finite off-site terms in the expression of the forces arise when using ultrasoft pseudopotentials. However this case is not explicitly treated in this chapter."} \footnote{We
    ve left in Cococcioni's final comment on ultrasoft PPs, because this may be important when discussing recent studies that attempt to calculate $U$ at low cost using USPPs.}
\end{blockquote}

\subsection{Bond forces for the one-dimensional binary lattice}

Next, let us consider a simplified Hubbard chain of alternating $\alpha$ and $\beta$ species, as illustrated in Figure \ref{fig:chain_uv}. For the sake of simplicity, we are only considering next-nearest $V^{\alpha \beta}$ inter-site terms, but the following derivation can be easily extended to systems of higher complexity.
\begin{figure}[H]
    \centering
    \includegraphics[width=0.95\linewidth]{si_images/chainUV.pdf}
    \caption{One-dimensional Hubbard chain of alternating $\alpha$ and $\beta$ species. Only NN $V^{\alpha \beta}$ inter-site terms are considered for simplicity. The goal here is to provide a simplified picture of a binary TM-O system.}
    \label{fig:chain_uv}
\end{figure}
\noindent We are interested in the force acting on an $\alpha$-$\beta$ bond, and therefore along the inter-site vector $\bm \tau_{\alpha \beta} =  \bm \tau_{\alpha} - \bm \tau_{\beta}$, where $F_i \left( \bm \tau_{\alpha} - \bm \tau_{\beta} \right)$ corresponds to the $i$ direction of the inter-site bond force between $\alpha$ and $\beta$ \footnote{Here, we use $\nabla_{\bm u} f = \nabla f \cdot \bm u$, the expression of the directional gradient of a multivariate function $f$ in the direction of $\bm u$, in terms of the gradient $\nabla f$.}, 
\begin{align}
    F^{UV}_i \left( \bm \tau_{\alpha} - \bm \tau_{\beta} \right)
    = - \frac{\partial E_{UV}}{\partial \left(\tau_{\alpha i} - \tau_{\beta i} \right)}
    &= - \frac{\partial E_{UV}}{\partial \tau_{\alpha i}} + \frac{\partial E_{UV}}{\partial \tau_{\beta i}} 
    = F^{UV}_{\alpha,i} - F^{UV}_{\beta,i} \nonumber \\
    &= - \frac{\partial E_{U}}{\partial \tau_{\alpha i}} + \frac{\partial E_{U}}{\partial \tau_{\beta i}}
    - \frac{\partial E_{V}}{\partial \tau_{\alpha i}} + \frac{\partial E_{V}}{\partial \tau_{\beta i}} \label{eq:bond_force_a}
\end{align}
The individual contributions to these forces can be expressed as follows, calling on the derivatives that we have expressed in Eqn.'s \ref{eq:eu_deriv}, \ref{eq:ev_deriv}, and \ref{eq:occ_deriv}:
\begin{align}
    F^{U}_{\alpha,i} &=- \sum_{m,m',\sigma}
    2 \Re \left\{ u^{\alpha \alpha \sigma}_{mm'} \left(\tilde n^{\alpha \alpha \sigma}_{mm'} \right)_i \right\} \label{eq:Fu_a} \\ 
    F^{V}_{\alpha,i} &= - \sum_{m,m',\sigma} \left[ 
    v^{\alpha \beta \sigma}_{mm'} \left(\tilde n^{\alpha \beta \sigma}_{mm'} \right)_i + 
    v^{\beta \alpha \sigma}_{mm'} \left(\tilde n^{\alpha \beta \sigma}_{mm'} \right)^*_i \right] \nonumber \\
    &= - \sum_{m,m',\sigma} v^{\alpha \beta \sigma}_{mm'} \left[ 
     \left(\tilde n^{\alpha \beta \sigma}_{mm'} \right)_i + \left(\tilde n^{\alpha \beta \sigma}_{mm'} \right)^*_i \right] \nonumber \\
    &= - \sum_{m,m',\sigma} 2 \Re \left\{ v^{\alpha \beta \sigma}_{mm'} \left(\tilde n^{\alpha \beta \sigma}_{mm'} \right)_i \right\} \label{eq:Fv_a}
\end{align}
We have used the symmetry of the occupation numbers that can be shown following from Equation \ref{eq:nocc_defn},
\begin{align}
    n^{\gamma{\gamma'}\sigma}_{mm'} 
    &= \sum_{k,v} f^\sigma_{kv} \braket{\psi^\sigma_{kv}|\phi^{\gamma'}_{m'}} \braket{\phi^\gamma_{m}|\psi^\sigma_{kv}} 
    = \sum_{k,v} f^\sigma_{kv} \braket{\psi^\sigma_{kv}|\phi^\gamma_{m}}^* \braket{\phi^{\gamma'}_{m'}|\psi^\sigma_{kv}}^* \nonumber \\
    &= \left\{ \sum_{k,v} f^\sigma_{kv} \braket{\psi^\sigma_{kv}|\phi^\gamma_{m}} \braket{\phi^{\gamma'}_{m'}|\psi^\sigma_{kv}} \right\}^* 
    = \left\{ n^{{\gamma'}\gamma\sigma}_{m'm} \right\}^* = n^{{\gamma'}\gamma\sigma}_{m'm},
\end{align}
where the last step was justified on the grounds that occupation numbers must be real. This symmetry holds for $v^{\gamma{\gamma'}\sigma}_{mm'}$, in accordance with Equation \ref{eq:ev_deriv}, which permits the factorization in Equations \ref{eq:Fv_a} \& \ref{eq:Fv_inter}.

Next, we can write the bond force terms as follows, 
\begin{align}
    F^{U}_i \left( \bm \tau_{\alpha} - \bm \tau_{\beta} \right) = F^{U}_{\alpha,i} - F^{U}_{\beta,i} &= - 2 \Re \left\{ \sum_{m,m',\sigma} \left[ 
    u^{\alpha \alpha \sigma}_{mm'} \left(\tilde n^{\alpha \alpha \sigma}_{mm'} \right)_i -
    u^{\beta \beta \sigma}_{mm'} \left(\tilde n^{\beta \beta \sigma}_{mm'} \right)_i
    \right] \right\} \\ 
    F^{V}_i \left( \bm \tau_{\alpha} - \bm \tau_{\beta} \right) = F^{V}_{\alpha,i} - F^{V}_{\beta,i} &= - 2 \Re \left\{ \sum_{m,m',\sigma} \left[ 
    v^{\alpha \beta \sigma}_{mm'} \left(\tilde n^{\alpha \beta \sigma}_{mm'} \right)_i - 
    v^{\beta \alpha \sigma}_{mm'} \left(\tilde n^{\beta \alpha \sigma}_{mm'} \right)_i \right] \right\} \nonumber \\
    &= - 2 \Re \left\{ \sum_{m,m',\sigma}  
    v^{\alpha \beta \sigma}_{mm'} \left[ 
    \left(\tilde n^{\alpha \beta \sigma}_{mm'} \right)_i - 
    \left(\tilde n^{\beta \alpha \sigma}_{mm'} \right)_i \right] \right\} \label{eq:Fv_inter}
\end{align} 
\footnote{The reader may notice that in $F^{V}_{i}$ terms, the contributions from $\alpha'$ and $\beta'$ sites are neglected, as shown in Figure \ref{fig:chain_uv}. These would intuitively ``tug on" the corresponding $\beta$ and $\alpha$ sites in the opposite direction.} At this point, it is helpful to keep in mind that $u^{\alpha \alpha \sigma}_{mm'} \ \propto \ U^{\alpha}$, $u^{\beta \beta \sigma}_{mm'} \ \propto \ U^{\beta}$, and $v^{\alpha \beta \sigma}_{mm'} \ \propto \ V^{\alpha \beta}$ from Eqs. \ref{eq:eu_deriv} \& \ref{eq:ev_deriv}. We see that while the contribution to the force from $V^{\alpha \beta}$ is only dependent on the difference between $\tilde n$ occupation number derivatives, the contribution from $U$ values depends additionally on the relative magnitude between $U^{\alpha}$ and $U^{\beta}$. 

This is not to say that the former is insignificant, but rather that both +$U$ (to both species) and +$V$ contributions should be accounted for if one is interested in correcting for $\alpha$-$\beta$ bond length. It is common practice to add a correction to $d$ and $f$-block species alone (``$\alpha$" sites). Adding a $U$ value to O-$p$ sites (``$\beta$"), if large enough, could possibly reverse the sign of $F^{U}_i \left( \bm \tau_{\alpha} - \bm \tau_{\beta} \right)$. This could explain the improved bond lengths after adding +$U$ corrections to O-$p$ sites, and not solely to TM-$d$ and $f$ sites.

Based on TMO test cases in this study, and the results of other studies \cite{LinscottEtAl, campo_jr_extended_2010}, it appears that applying a $U$ to TM sites only overestimates TM-O bond-length from DFT. Furthermore, both (a) adding a +U to the O-$p$ manifold \cite{LinscottEtAl} and (b) +V between TM and O sites \cite{campo_jr_extended_2010} appears to reduce the severity of this overestimation. This leads us to assume that starting from the DFT (no +$U$/$J$/$V$) optimized structure, $F^U_\alpha < 0$ (tensile) and $F^U_\beta < 0$ (compressive), where it's still the case that $|F^U_\alpha| > |F^U_\beta|$ (tensile), because the bond lengths are still overestimated with +U(O-p). And finally, $F^V_\alpha - F^V_\beta > 0$ (compressive), because Campo et al. appear have found that +$V$ helped correct for TM-O bond length in NiO \cite{campo_jr_extended_2010}.

\printbibliography